\documentclass{article}
\pdfoutput=1
\usepackage{fullpage}
\usepackage{textcomp}
\usepackage{subcaption}
\usepackage{enumerate}
\usepackage{amsmath}
\usepackage{amsfonts}
\usepackage{graphicx}

\newtheorem{lemma}{Lemma}
\newtheorem{theorem}[lemma]{Theorem}
\newtheorem{corollary}[lemma]{Corollary}
\newtheorem{definition}[lemma]{Definition}

\graphicspath{{./figs/}}

\title{Stochastic closest-pair problem and most-likely nearest-neighbor search in tree spaces}
\author{
    Jie Xue\footnote{Dept. of Computer Science and Engg., Univ. of Minnesota --- Twin Cities, 4-192 Keller Hall, 200 Union St. SE, Minneapolis, MN 55455, USA}\\
    \texttt{xuexx193@umn.edu}
    \and
    Yuan Li\footnotemark[1]\\
    \texttt{lixx2100@umn.edu}
}
\date{}

\begin{document}

\maketitle
\begin{abstract}
Let $\mathcal{T}$ be a tree space (or tree network) represented by a
    weighted tree with $t$ vertices, and $S$ be a set of $n$ stochastic points
    in $\mathcal{T}$, each of which has a fixed location with an independent
    existence probability.
We investigate two fundamental problems under such a stochastic setting, the
    closest-pair problem and the nearest-neighbor search.
For the former, we study the computation of the $\ell$-threshold probability
    and the expectation of the closest-pair distance of a realization of $S$.
We propose the first algorithm to compute the $\ell$-threshold probability in
    $O(t+n \log n+ \min\{tn,n^2\})$ time for any given threshold $\ell$, which
    immediately results in an $O(t+ \min\{tn^3,n^4\})$-time algorithm for
    computing the expected closest-pair distance.
Based on this, we further show that one can compute a
    $(1+\varepsilon)$-approximation for the expected closest-pair distance in
    $O(t+ \varepsilon^{-1}\min\{tn^2,n^3\})$ time, by arguing that the expected
    closest-pair distance can be approximated via $O(\varepsilon^{-1} n)$
    threshold probability queries.
For the latter, we study the $k$ most-likely nearest-neighbor search ($k$-LNN)
    via a notion called $k$ most-likely Voronoi Diagram ($k$-LVD).
We show that the size of the $k$-LVD $\varPsi_\mathcal{T}^S$ of $S$ on
    $\mathcal{T}$ is bounded by $O(kn)$ if the existence probabilities of the
    points in $S$ are constant-far from 0.
Furthermore, we establish an $O(kn)$ average-case upper bound for the size of
    $\varPsi_\mathcal{T}^S$, by regarding the existence probabilities as i.i.d.
    random variables drawn from some fixed distribution.
Our results imply the existence of an LVD data structure which answers $k$-LNN
    queries in $O(\log n + k)$ time using average-case $O(t+k^2 n)$ space, and
    worst-case $O(t+k n^2)$ space if the existence probabilities are
    constant-far from 0.
Finally, we also give an $O(t+ n^2 \log n + n^2 k)$-time algorithm to construct
    the LVD data structure.
\end{abstract}

\section{Introduction}
In many real-world applications, due to the existence of noise or limitation of
    devices, the data obtained may be imprecise or not totally reliable.
In this situation, certain datasets may fail to well capture the features of
    data and uncertain ones are more preferable.
Motivated by this, the topic of uncertain data has received significant
    attentions in the last decades.
Many classical problems have been investigated under uncertainty, including
    convex hull, minimum spanning tree, range search, linear separability and
    so forth \cite{
        agarwal2013nearest,
        agarwal2014convex,
        kamousi2011stochastic2,
        suri2014most,
        suri2013most}. 
Among these works, there are two kinds of commonly used models of uncertainty:
    existential uncertainty and locational uncertainty.
In the former, each (stochastic) data point has a fixed location with an
    uncertain existence depicted by an independent existence probability, while
    in the latter the location of each point is uncertain and described as a
    distribution.

The closest-pair problem and nearest-neighbor search are two interrelated
    fundamental problems, which have numerous applications in various areas.
The uncertain versions of both the problems have also been studied recently
    \cite{
        agarwal2013nearest,
        huang2015approximating,
        kamousi2014closest,
        suri2014most}. 
Let $S$ be a set of $n$ stochastic points in some metric space $\mathcal{X}$.
Concerning the closest-pair under uncertainty, a basic question one may ask is
    how to compute elementary statistics about the stochastic closest-pair of
    $S$, e.g., the probability that the closest-pair distance of a realization
    of $S$ is at least $\ell$, the expected closest-pair distance, etc.
Unfortunately, most problems of this kind have been shown to be NP-hard or
    \#P-hard for general metric, and some of them remain \#P-hard even when
    $\mathcal{X} = \mathbb{R}^d$ for $d \geq 2$
    \cite{huang2015approximating,kamousi2014closest}.
Due to the hardness of the stochastic closest-pair problems in general and
    Euclidean space, it is then natural to ask whether these problems are
    easier in other kinds of metric spaces such as tree space (or tree network).
Concerning the nearest-neighbor search under uncertainty, an important problem
    is the most-likely nearest-neighbor (LNN) search \cite{suri2014most}, which
    looks for the data point in $S$ with the greatest probability of being the
    nearest-neighbor of a query point $q$.
The LNN search induces the concept of most-likely Voronoi diagram (LVD), which
    decomposes $\mathcal{X}$ into connected cells such that the query points in
    the same cell have the same LNN.
Some results about the LVD and LNN search are given in \cite{suri2014most}.
However, the existing results are only for 1-LNN search in
    $\mathcal{X} = \mathbb{R}^1$.
More generally, one may consider the problem of $k$-LNN search, which reports
    the $k$ points in $S$ with the greatest probabilities of being the
    nearest-neighbor of $q$.
Furthermore, it is also interesting to investigate the LNN search and LVD in
    less-trivial non-Euclidean metric spaces.

\begin{figure}[h]
    \centering
    \includegraphics[height=4cm]{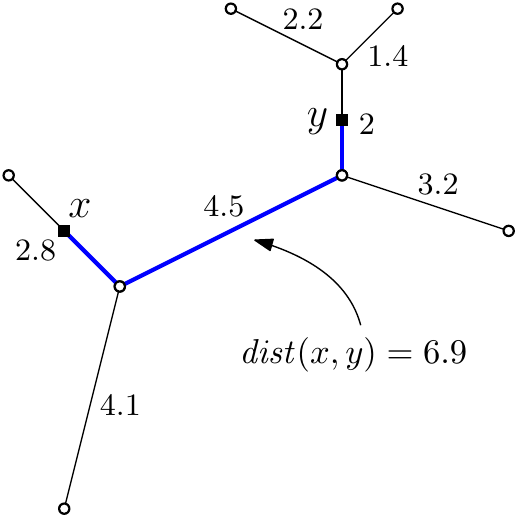}
    \caption{An example of tree space.}
    \label{fig:1}
\end{figure}

With the above motivations, in this paper, we study the stochastic closest-pair
    (SCP) problem and $k$ most-likely nearest-neighbor ($k$-LNN) search in tree
    spaces.
A \textit{tree space} $\mathcal{T}$ is represented by a positively-weighted
    tree $T$ where the weight of each edge depicts its ``length''.
Formally, $\mathcal{T}$ is the geometric realization of $T$, in which each edge
    weighted by $w$ is isometric to the interval $[0,w]$.
There is a natural metric over $\mathcal{T}$ which defines the distance
    $\mathit{dist}(x,y)$ as the length of the (unique) simple path between $x$
    and $y$ in $\mathcal{T}$.
See Figure~\ref{fig:1} for an example of tree space.
Following \cite{huang2015approximating,kamousi2014closest,suri2014most}, we
    study the problems under existential uncertainty: each stochastic point has
    a fixed location (in $\mathcal{T}$) associated with an (independent)
    existence probability.
Due to limited space, the proofs of \textbf{all} lemmas and some theorems are
    deferred to Appendix~\ref{prfs}.
\smallskip

\noindent
\textbf{Our result.}
Let $\mathcal{T}$ be a tree space represented by a $t$-vertex weighted tree
    $T$, and $S$ be the given set of $n$ stochastic points in $\mathcal{T}$
    each of which is associated with an existence probability.
A \textit{realization} of $S$ refers to a random sample of $S$ in which each
    point is sampled with its existence probability.

For the SCP problem, define $\kappa(S)$ as a random variable indicating the
    closest-pair distance of a realization of $S$.
We first show that the $\ell$-threshold probability of $\kappa(S)$ (i.e., the
    probability that $\kappa(S)$ is at least $\ell$) can be computed in
    $O(t+n \log n+ \min\{tn,n^2\})$ time for any given positive threshold
    $\ell$.
Based on this, we immediately obtain an $O(t+ \min\{tn^3,n^4\})$-time algorithm
    for computing the expected closest-pair distance, i.e., the expectation of
    $\kappa(S)$.
We then further show that one can approximate the expected closest-pair
    distance within a factor of $(1+\varepsilon)$ in
    $O(t+ \varepsilon^{-1}\min\{tn^2,n^3\})$ time, by arguing that the expected
    closest-pair distance can be approximated via $O(\varepsilon^{-1} n)$
    threshold probability queries.

For the LNN search, we first study the size of the the $k$-LVD
    $\varPsi_\mathcal{T}^S$ of $S$ on $\mathcal{T}$.
A matching $O(n^2)$ upper bound for the worst-case size of
    $\varPsi_\mathcal{T}^S$ is given.
More interestingly, we show that (1) the worst-case size of
    $\varPsi_\mathcal{T}^S$ is $O(kn)$, if the existence probabilities of the
    points in $S$ are constant-far from 0; (2) the average-case size of
    $\varPsi_\mathcal{T}^S$ is $O(kn)$, if the existence probabilities are
    i.i.d. random variables drawn from a fixed distribution.
These results further imply the existence of an LVD data structure which
    answers $k$-LNN queries in $O(\log n + k)$ time using average-case
    $O(t+k^2 n)$ space, and worst-case $O(t+k^2 n)$ space if the existence
    probabilities of the points are constant-far from 0.
Finally, we give an $O(t+ n^2 \log n + n^2 k)$-time algorithm to construct such
    a data structure.
\smallskip

\noindent
\textbf{Related work.}
The topic of uncertain data has received significant attentions in various
    areas such as computational geometry, algorithms, databases, etc.
Many classical problems have been studied in stochastic settings, including
    convex hull \cite{agarwal2014convex,loffler2010largest,suri2013most},
    minimum spanning tree \cite{kamousi2011stochastic2},
    range search \cite{agarwal2012range,agarwal2016range},
    linear separability \cite{fink2016hyperplane,xue2016separability},
    top-$k$ queries \cite{chen2013efficient,ge2009top}, etc.

More relevantly, the stochastic versions of the closest-pair problem and
    nearest-neighbor search have also been investigated in
    \cite{
        agarwal2013nearest,
        agarwal2012nearest,
        huang2015approximating,
        kamousi2014closest,suri2014most}.
Kamousi et al. \cite{kamousi2014closest} show that computing the
    $\ell$-threshold probability of the closest-pair distance and some variants
    of the problem are \#P-hard under existential uncertainty even in
    $\mathbb{R}^2$.
The nearest-neighbor search is also considered in \cite{kamousi2014closest}
    under existential uncertainty, but the studied problem is to find the point
    minimizing the expected distance to the query point instead of the LNN.
Huang et al. \cite{huang2015approximating} give hardness results and randomized
    approximation algorithms for some stochastic closest-pair related problems
    under general metric.
It is shown in \cite{huang2015approximating} that computing the expected
    closest-pair distance under existential uncertainty is \#P-hard in a
    general metric space.
Agarwal et al. \cite{agarwal2013nearest,agarwal2012nearest} study the uncertain
    nearest-neighbor search, but their main focus is the locational uncertainty
    and the problems studied are quite different from the LNN search.
Suri et al. \cite{suri2014most} investigate the LNN search and give upper
    bounds for the complexity of the LVD as well as the way to construct the
    LVD.
    
However, only the case of 1-LNN search in $\mathbb{R}^1$ is studied in
    \cite{suri2014most}.
The problem in general Euclidean space and non-Euclidean metric spaces is quite
    open, so is the $k$-LNN search.

\section{The stochastic closest-pair problems}
Let $\mathcal{T}$ be a tree space represented by a $t$-vertex weighted tree $T$ and $S = \{a_1,\dots,a_n\} \subset \mathcal{T}$ be a set of stochastic points where $a_i$ has an existence probability $\pi_{a_i}$.
We use $\kappa(S)$ to denote the random variable indicating the closest-pair distance of a realization of $S$ (if the realization is of size less than 2, we simply set its closest-pair distance to be 0).

\subsection{Computing the threshold probability} \label{thpr}
We study the problem of computing the probability that $\kappa(S)$ is at least $\ell$ for a given threshold $\ell$.
We call this quantity the $\ell$-\textit{threshold probability} or simply \textit{threshold probability} of $\kappa(S)$, and denote it by $C_{\geq \ell}(S)$.
We show that $C_{\geq \ell}(S)$ can be computed in $O(t+n \log n+ \min\{tn,n^2\})$ time.
This result gives us an $O(t+n^2)$ upper bound for $t = \Omega(n)$ and an $O(n \log n + tn)$ bound for $t = O(n)$.
In the rest of this section, we first present an $O(t+n^3)$-time algorithm for computing $C_{\geq \ell}(S)$, and then show how to improve it to achieve the desired bound.
For simplicity of exposition, we assume $a_1,\dots,a_n$ have distinct locations in $\mathcal{T}$ (note that the degenerate case can be easily handled by replacing the stochastic points at the same location with a new stochastic point with an appropriate existence probability).

\subsubsection{An \texorpdfstring{$O(t+n^3)$}{t+n3}-time algorithm} \label{sec_nqtime}
In order to conveniently and efficiently handle the stochastic points in a tree space, we begin with a preprocessing, which reduce the problem to a more regular setting.
\begin{theorem} \label{preproc}
    Given $\mathcal{T}$ and $S$, one can compute in $O(t+n \log n)$ time a new tree space $\mathcal{T}' \subseteq \mathcal{T}$ represented by an $O(n)$-vertex weighted tree $T'$ such that $S \subset \mathcal{T}'$ and every point in $S$ is located at some vertex of $T'$.
    \textnormal{(See Appendix~\ref{prfpreproc} for a proof.)}
\end{theorem}
By the above theorem, we use $O(t+n \log n)$ time to compute such a new tree space.
Using this tree space as well as the $O(n)$-vertex tree representing it, the problem becomes more regular: every stochastic point in $S$ is located at a vertex.
We can further make the stochastic points one-to-one corresponding to the vertices by adding dummy points with existence probability 0 to the ``empty'' vertices.
In such a regular setting, we then consider how to compute the $\ell$-threshold probability.
For convenience, we still use $T$ to denote the representation of the (new) tree space and $S = \{a_1,\dots,a_n\}$ the stochastic dataset (though the actual size of $S$ may be larger than $n$ due to the additional dummy points, it is still bounded by $O(n)$).
Since the vertices of $T$ are now one-to-one corresponding to the points in $S$, we also use $a_i$ to denote the corresponding vertex of $T$.

\begin{figure}[htpb]
    \centering
    \includegraphics[height=4cm]{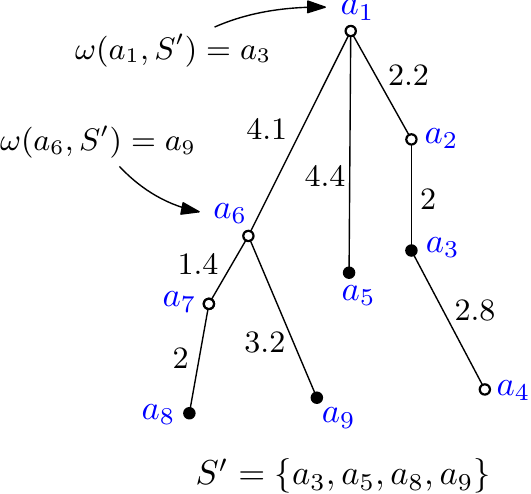}
    \caption{An illustration of witness.}
    \label{fig:2}
\end{figure}

As we are working on a tree space, a natural idea for solving the problem is to exploit the recursive structure of the tree and to compute $C_{\geq \ell}(S)$ in a recursive fashion.
To this end, we need to define an important concept called \textit{witness}.
We make $T$ rooted by setting $a_1$ as its root.
The subtree rooted at a vertex $x$ is denoted by $T_x$.
Also, we use $V(T_x)$ to denote the set of the stochastic points lying in $T_x$, or equivalently, the set of the vertices of $T_x$.
The notations $\bar{p}(x)$ and $\text{ch}(x)$ are used to denote the parent of $x$ and the set of the children of $x$, respectively (for convenience we set $\bar{p}(a_1) = a_1$).
\begin{definition}
    Let $\mathit{dep}(a_i)$ be the depth of $a_i$ in $T$, i.e., $\mathit{dep}(a_i) = \mathit{dist}(a_1,a_i)$.
    For any $a_i$ and $a_j$, we define $a_i \prec a_j$ if $\mathit{dep}(a_i)<\mathit{dep}(a_j)$, or $\mathit{dep}(a_i)=\mathit{dep}(a_j)$ and $i<j$.
    Clearly, the relation $\prec$ is a strict total order over $S$ (also, over the vertices of $T$).
    For any subset $S' \subseteq S$ and any vertex $a_i$ of $T$, we define the \textbf{witness} of $a_i$ with respect to $S'$, denoted by $\omega(a_i,S')$, as the smallest vertex in $V(T_{a_i}) \cap S'$ under the $\prec$-order.
    If $V(T_{a_i}) \cap S' = \emptyset$, we say $\omega(a_i,S')$ is not defined.
\end{definition}
See Figure~\ref{fig:2} for an illustration of witness.
We say a subset $S' \subseteq S$ \textit{legal} if the closest-pair distance of $S'$ is at least $\ell$.
The following lemma allows us to verify the legality of a subset by using the witnesses, which will be used later.
\begin{lemma} \label{th1}
    For any $S' \subseteq S$, we have $S'$ is legal if and only if any point $a_i \in S\backslash\{a_1\}$ satisfies one of the following three conditions: \\
    \textnormal{(1)} $\omega(a_i,S')$ is not defined; \\
    \textnormal{(2)} $\omega(a_i,S') = \omega(\bar{p}(a_i),S')$; \\
    \textnormal{(3)} $\mathit{dist}(\omega(a_i,S'),\omega(\bar{p}(a_i),S')) \geq \ell$. \\
    If $a_i$ satisfies one of the above conditions, we say that $S'$ is \textbf{locally legal} at $a_i$.
\end{lemma}
In order to compute $C_{\geq \ell}(S)$, we define, for all $x \in S$ and $y \in V(T_{\bar{p}(x)})$,
\begin{equation*}
P_y(x) = \left\{
\begin{array}{ll}
\Pr\limits_{S' \subseteq_\text{R} V(T_x)} \left[ S' \text{ is legal and } \omega(x,S')=y \right] & \text{if } y \in V(T_x), \\[13pt]
\Pr\limits_{S' \subseteq_\text{R} V(T_x)} \left[ S' \cup \{y\} \text{ is legal and } \omega(\bar{p}(x),S' \cup \{y\})=y \right] & \text{if } y \in V(T_{\bar{p}(x)}) \backslash V(T_x).
\end{array}
\right.
\end{equation*}
Here the notation $\subseteq_\text{R}$ means that the former is a realization of the latter, i.e., a random sample obtained by sampling each point with its existence probability.
With the above, we immediately have that $C_{\geq \ell}(S) = \sum_{i=1}^{n} P_{a_i}(a_1) - P_0$, where $P_0$ is the probability that a realization of $S$ contains exactly one point.
We then show how $P_y(x)$ can be computed in a recursive way.
\begin{theorem} \label{comp1}
    For $x \in S$ and $y \in V(T_x)$, we have that
    \begin{equation*}
    P_y(x) = Q \cdot \prod_{c \in \textnormal{ch}(x)} P_y(c),
    \end{equation*}
    where $Q = \pi_x$ if $x=y$ and $Q = 1-\pi_x$ if $x \neq y$.
\end{theorem}
\textit{Proof.}
By definition, when $y \in V(T_x)$, $P_y(x)$ is the probability that a realization $S' \subseteq_\text{R} V(T_x)$ is legal and $\omega(x,S')=y$.
If $x = y$, $x$ must be in $S'$ in order to have $\omega(x,S')=y$.
Otherwise, if $x \neq y$, $x$ must not be in $S'$.
Thus, the meaning of the factor $Q$ in the formula is clear.
Then we consider the vertices in $V(T_x)$ other than $x$.
Clearly, if $S'$ is legal, then $S' \cap V(T_c)$ is also legal for any $c \in \text{ch}(x)$.
Also, if $\omega(x,S')=y$, then $w(c,S' \cap V(T_c)) = y$ if $y \in V(T_c)$ and $w(\bar{p}(c),(S' \cap V(T_c)) \cup \{y\}) = y$ if $y \notin V(T_c)$.
Therefore, the probabilities of all the legal instances $S' \subseteq V(T_x)$ satisfying $\omega(x,S')=y$ are counted by the right-hand side of the formula.
It suffices to show that the right-hand side does not overestimate the probability, i.e., every instance $S'$ counted by the right-hand side truly satisfies the desired properties: $S'$ is legal and $\omega(x,S')=y$.
Let $S'$ be an instance counted by the right-hand side.
The property $\omega(x,S')=y$ is obviously satisfied.
To see $S'$ is legal, by Lemma~\ref{th1}, we only need to verify the local legality of $S'$ at every vertex in $S \backslash \{a_1\}$.
Since $S'$ does not contain any vertices outside $V(T_x)$, the local legalities at $x$ and all $a_i \notin V(T_x)$ clearly hold.
Also, $S'$ is locally legal at any $a_i \in V(T_x) \backslash (\text{ch}(x) \cup \{x\})$, because each factor $P_y(c)$ forces $S' \cap V(T_c)$ to be legal.
Now we verify that $S'$ is locally legal at any $c \in \text{ch}(x)$.
If $y \in V(T_c)$, then $\omega(c,S') = \omega(x,S') = y$ and hence $S'$ is legal at $c$.
If $y \notin V(T_c)$, then the factor $P_y(c)$ forces $(S' \cap V(T_c)) \cup \{y\}$ to be legal and thus either $\omega(c,S')$ is not defined or $\mathit{dist}(\omega(c,S'),y) \geq \ell$, which implies that $S'$ is legal at $c$.
\hfill $\Box$
\begin{theorem} \label{comp2}
    For $x \in S$ and $y \in V(T_{\bar{p}(x)}) \backslash V(T_x)$, we have that
    \begin{equation*}
    P_y(x) = \prod_{a_i \in V(T_x)} (1-\pi_{a_i})  + \sum_{z \in \varGamma} P_z(x),
    \end{equation*}
    where $\varGamma = \{z \in V(T_x): y \prec z \text{ and } \mathit{dist}(z,y) \geq \ell \}$.
    \textnormal{(See Appendix~\ref{prfcomp2} for a proof.)}
\end{theorem}
By the above two theorems, the values of all $P_y(x)$ can be computed as follows.
We enumerate $x \in S$ from the greatest to the smallest under $\prec$-order.
For each $x$, we first compute all $P_y(x)$ for $y \in V(T_x)$ by applying Theorem~\ref{comp1}.
After this, we are able to compute all $P_y(x)$ for $y \in V(T_{\bar{p}(x)}) \backslash V(T_x)$ by applying Theorem~\ref{comp2}.
The entire process takes $O(n^3)$ time.
Once we have the values of all $P_y(x)$, $C_{\geq \ell}(S)$ can be computed straightforwardly.
Including the time for preprocessing, this gives us an $O(t+n^3)$-time algorithm for computing $C_{\geq \ell}(S)$.

\subsubsection{Improving the runtime}
We first show how to improve the runtime of the above algorithm to $O(t+n^2)$.
Note that computing all $P_y(x)$ for $x \in S$ and $y \in V(T_x)$ takes only $O(n^2)$ time in total, as we can charge the time for computing $P_y(x)$ to the pairs $(y,c)$ for $c \in \text{ch}(x)$ and thus each pair of vertices is charged at most a constant amount of time.
So the bottleneck is the computation of $P_y(x)$ for $y \in V(T_{\bar{p}(x)}) \backslash V(T_x)$.
For a specific $x \in S$, we want to compute all $P_y(x)$ for $y \in V(T_{\bar{p}(x)}) \backslash V(T_x)$ in linear time.
To achieve this, we review the formula given in Theorem~\ref{comp2}.
Assume that $V(T_x) = \{z_1,\dots,z_m\}$ where $z_1 \prec \cdots \prec z_m$, and $V(T_{\bar{p}(x)}) \backslash V(T_x) = \{y_1,\dots,y_r\}$ where $y_1 \prec \cdots \prec y_r$.
Define
\begin{equation*}
\varGamma_{y_i} = \{z \in V(T_x): y_i \prec z \text{ and } \mathit{dist}(z,y_i) \geq \ell \}
\end{equation*}
for $i \in \{1,\dots,r\}$.
Then $P_{y_i}(x)$ is just the sum of $\prod_{j=1}^{m} (1-\pi_{z_j})$ and all $P_z(x)$ for $z \in \varGamma_{y_i}$.
\begin{theorem} \label{th4}
    Each set $\varGamma_{y_i}$ is a suffix of the sequence $(z_1,\dots,z_m)$, i.e., $\varGamma_{y_i} = \{z_j,z_{j+1},\dots,z_m\}$ for some $j \in \{1,\dots,m\}$.
    Furthermore, we have that $\varGamma_{y_1} \subseteq \cdots \subseteq \varGamma_{y_k} \supseteq \cdots \supseteq \varGamma_{y_r}$ for some $k \in \{1,\dots,t\}$.
    \textnormal{(See Appendix~\ref{prfth4} for a proof.)}
\end{theorem}
The above observation gives us the idea to efficiently compute the values of $P_{y_1}(x),\dots,P_{y_r}(x)$.
Instead of computing $P_{y_i}(x)$ straightforwardly using the formula given in Theorem~\ref{comp2}, we compute each $P_{y_i}(x)$ by modifying $P_{y_{i-1}}(x)$.
Specifically, we first compute $P_{y_1}(x)$ straightforwardly and then begin to compute $P_{y_2}(x),\dots,P_{y_r}(x)$ in order.
If $\varGamma_{y_i} \subseteq \varGamma_{y_{i-1}}$, we compute $P_{y_i}(x)$ by subtracting all $P_z(x)$ for $z \in \varGamma_{y_{i-1}} \backslash \varGamma_{y_i}$ from $P_{y_{i-1}}(x)$.
Otherwise, if $\varGamma_{y_i} \supseteq \varGamma_{y_{i-1}}$, we compute $P_{y_i}(x)$ by adding all $P_z(x)$ for $z \in \varGamma_{y_i} \backslash \varGamma_{y_{i-1}}$ to $P_{y_{i-1}}(x)$.
According to Theorem~\ref{th4}, in the entire process, each $P_z(x)$ for $z \in \{z_1,\dots,z_m\}$ is at most added and subtracted once.
Therefore, with the sequence $(z_1,\dots,z_m)$ in hand, it is easy to compute $P_{y_1}(x),\dots,P_{y_r}(x)$ in $O(n)$ time.
Note that the sequence $(z_1,\dots,z_m)$ can be easily obtained in $O(n)$ time, if we sort all the points $a_1,\dots,a_n$ in $\prec$-order at the beginning of the algorithm.
This improves the overall time complexity to $O(t+n^2)$.

Indeed, we can further improve the runtime to $O(t+n \log n +\min\{tn,n^2\})$.
In other words, we show that $C_{\geq \ell}(S)$ can be computed in $O(n \log n + tn)$ time when $t = O(n)$.
To achieve this, we retrospect the original tree space (before the preprocessing) which is represented by a $t$-vertex tree.
Intuitively, if $t$ is significantly smaller than $n$, then most stochastic points in $S$ are located inside the interiors of the edges of the original tree.
In this case, after the preprocessing, we will have a lot of ``chain'' structures in the new tree $T$.
This gives us the insight to further improve our algorithm.

\begin{figure}[htpb]
    \centering
    \includegraphics[height=4cm]{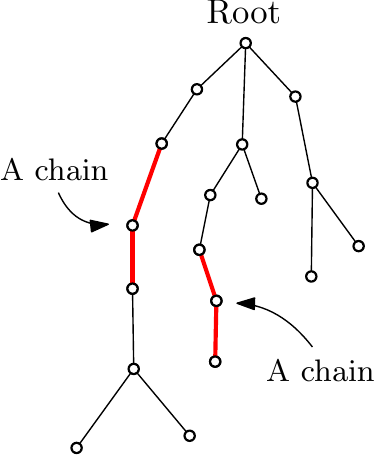}
    \caption{An example of chains.}
    \label{fig:3}
\end{figure}

\begin{definition}
    A \textbf{chain} of $T$ is a sequence of vertices $(b_1,\dots,b_k)$ satisfying \\
    \textnormal{(1)} $b_i$ is the only child of $b_{i-1}$ for $i \in \{2,\dots,k\}$; \\
    \textnormal{(2)} $b_k$ has at most one child; \\
    \textnormal{(3)} $b_1$ is either the root or the only child of $\bar{p}(b_1)$. \\
    \textnormal{(See Figure~\ref{fig:3} for an example of chain.)}
    A chain is \textbf{maximal} if it is not properly contained in another chain.
    A vertex of $T$ is called \textbf{chain vertex} if it is contained in some chain.
    Otherwise, it is called \textbf{non-chain vertex}.
\end{definition}
\begin{theorem} \label{numofnonch}
    If $\mathcal{T}$ is a tree space represented by a $t$-vertex tree and $\mathcal{T}' \subseteq \mathcal{T}$ is also a tree space represented by a rooted tree $T$, then the number of the non-chain vertices of $T$ is $O(t)$.
    \textnormal{(See Appendix~\ref{prfnumofnonch} for a proof.)}
\end{theorem}
One can easily verify that when removing all the non-chain vertices (and their adjacent edges) from $T$, each connected component of the remaining forest corresponds to a maximal chain of $T$.
Thus, the number of the maximal chains of $T$ is also bounded by $O(t)$.

Now we explain why the chains of $T$ are helpful for us.
Let $(b_1,\dots,b_k)$ be a chain of $T$.
For convenience of exposition, we assume $b_k$ has a (unique) child $b_{k+1}$ and $b_1$ has the parent $b_0$.
Our previous algorithm takes $O(kn)$ time to compute all $P_y(x)$ for $x \in \{b_1,\dots,b_k\}$ and $y \in V(\bar{p}(x))$.
To improve the runtime, we want that these values can be computed in $O(n)$ time.
This seems impossible as the number of the values to be computed is $\Theta(kn)$ in worst case.
However, instead of computing these values explicitly, we can compute them implicitly.
Note that $P_y(b_i)$ is defined only when $y \in \{b_{i-1},\dots,b_k\} \cup V(T_{b_{k+1}})$.
Set $\sigma_0 = 1$ and $\sigma_i = \prod_{j=1}^{i} (1-\pi_{b_j})$ for $i \in \{1,\dots,k\}$.
Let $b_i \in \{b_1,\dots,b_k\}$ be a vertex in the chain.
By Theorem~\ref{comp1}, we observe the following.
First, for any $y \in V(T_{b_{k+1}})$, we have that $P_y(b_i) = P_y(b_{k+1}) \cdot \sigma_k / \sigma_{i-1}$.
Furthermore, we have that $P_{b_i}^{}(b_i) = \pi_{b_i} \cdot P_{b_i}^{}(b_{i+1})$ and 
\begin{equation*}
P_{b_j}^{}(b_i) = P_{b_j}^{}(b_j) \cdot \frac{\sigma_{j-1}}{\sigma_{i-1}} = P_{b_j}^{}(b_{j+1}) \cdot \frac{\pi_{b_j} \sigma_{j-1}}{\sigma_{i-1}}
\end{equation*}
for $j \in \{i+1,\dots,k\}$.
Thus, as long as we know the values of $\sigma_1,\dots,\sigma_k$ and $P_{b_0}^{}(b_1),\dots,P_{b_{k-1}}^{}(b_k)$, any $P_y(x)$ with $x \in \{b_1,\dots,b_k\}$ can be computed in constant time (note that the values of $P_y(b_{k+1})$ are already in hand when we deal with the chain).
In other words, to implicitly compute all $P_y(x)$ for $x \in \{b_1,\dots,b_k\}$, it suffices to compute $\sigma_1,\dots,\sigma_k$ and $P_{b_0}^{}(b_1),\dots,P_{b_{k-1}}^{}(b_k)$, and associate to each $b_i$ the values of $\sigma_i$ and $P_{b_{i-1}}^{}(b_i)$.
Clearly, one can easily compute $\sigma_1,\dots,\sigma_k$ in $O(k)$ time.
We then show that $P_{b_0}^{}(b_1),\dots,P_{b_{k-1}}^{}(b_k)$ can be computed in $O(n)$ time.
Define $A_i = \{z \in V(T_{b_i}):\mathit{dist}(z,b_{i-1}) \geq \ell\}$, then $A_k \subseteq A_{k-1} \subseteq \cdots \subseteq A_1$ and each $A_i$ is a suffix of the $\prec$-order sorted sequence of the vertices in $V(T_{b_0})$.
Now by Theorem~\ref{comp2}, one can deduce that
\begin{equation*}
P_{b_{i-1}}^{}(b_i) = (1 - \pi_{b_i}) \cdot P_{b_i}^{}(b_{i+1}) + \sum_{z \in A_i \backslash A_{i+1}} Q_z \cdot P_z(b_{i+1}),
\end{equation*}
where $Q_z = \pi_{b_i}$ if $z = b_i$ and $Q_z = 1 - \pi_{b_i}$ otherwise.
Thus, if the computation is taken in the order $P_{b_{k-1}}^{}(b_k),\dots,P_{b_0}^{}(b_1)$, then each $P_{b_{i-1}}^{}(b_i)$ can be easily computed in $O(|A_i \backslash A_{i+1}|)$ time.
In this way, we use $O(n)$ time to implicitly compute all $P_y(x)$ for $x \in \{b_1,\dots,b_k\}$.
It turns out that the computation task for any chain can be done in $O(n)$ time.

With this in hand, it is not difficult to compute all $P_y(x)$ in $O(tn)$ time.
We enumerate $x \in S$ from the greatest to the smallest under $\prec$-order.
For each $x$ visited, if $x$ is a non-chain vertex,  we use $O(n)$ time to explicitly compute all $P_y(x)$ in the previous way.
If $x$ is the deepest vertex of a chain, i.e., $x$ has no child or its child is a non-chain vertex, then we find the maximal chain containing $x$ and implicitly complete the computation task for this chain in $O(n)$ time.
Otherwise, if $x$ is a chain vertex but not the deepest one, we just skip it as all $P_y(x)$ have been implicitly computed previously.
The entire process takes $O(tn)$ time, as there are $O(t)$ non-chain vertices and $O(t)$ maximal chains.
Including the time for preprocessing and sorting $a_1,\dots,a_n$, we solve the problem in $O(n \log n + tn)$ time.
Combining with the case $t = \Omega(n)$, we finally conclude the following.
\begin{theorem}
    Given a weighted tree $T$ with $t$ vertices and a set $S$ of $n$ stochastic points in its tree space $\mathcal{T}$, one can compute the $\ell$-threshold probability of the closest-pair distance of $S$, $C_{\geq \ell}(S)$, in $O(t+n \log n +\min\{tn,n^2\})$ time.
\end{theorem}

\subsection{Computing the expected closest-pair distance} \label{ecdist}
Based on our algorithm for computing the threshold probability, we further study the problem of computing the expected closest-pair distance of $S$, i.e., the expectation of $\kappa(S)$.
It is easy to see that our algorithm in Section~\ref{thpr} immediately gives us an $O(t + \min\{tn^3,n^4\})$ algorithm to compute $\textbf{E}[\kappa(S)]$.
This is because the random variable $\kappa(S)$ has at most $\binom{n}{2}$ distinct possible values and hence we can compute $\textbf{E}[\kappa(S)]$ via $O(n^2)$ threshold probability ``queries'' with various thresholds $\ell$ (note that after preprocessing our algorithm answers each threshold probability query in $O(\min\{tn,n^2\})$ time).
\begin{theorem}
    Given a tree space $\mathcal{T}$ represented by a $t$-vertex weighted tree $T$ and a set $S$ of $n$ stochastic points in $\mathcal{T}$, one can compute the expected closest-pair distance of $S$, $\textnormal{\textbf{E}}[\kappa(S)]$, in $O(t + \min\{tn^3,n^4\})$ time.
\end{theorem}
If we want to compute the exact value of $\textbf{E}[\kappa(S)]$ (via threshold probability queries), $\Theta(n^2)$ queries are necessary in worst case.
So it is natural to ask whether we can use less queries to approximate $\textbf{E}[\kappa(S)]$.
In the rest of this section, we show that one can use $O(\varepsilon^{-1}n)$ threshold probability queries to achieve a $(1+\varepsilon)$-approximation for $\textbf{E}[\kappa(S)]$, which in turn gives us an $O(t + \varepsilon^{-1}\min\{tn^2,n^3\})$-time approximation algorithm for computing $\textbf{E}[\kappa(S)]$.

For simplicity of exposition, we assume that the stochastic points in $S$ are now one-to-one corresponding to the vertices of $T$ (this is what we have after preprocessing).
We begin with a simple case, in which the \textit{spread} of $T$, i.e., the ratio of the length of the longest edge to the length of the shortest edge is bounded by some polynomial of $n$.
In this case, to approximate $\textbf{E}[\kappa(S)]$ is fairly easy, and we only need $O(\varepsilon^{-1} \log n)$ threshold probability queries.
\begin{definition}
    For $\beta > \alpha > 0$ and $\tau>1$, the $(\alpha,\beta,\tau)$-\textbf{jump} is defined as the set
    \begin{equation*}
    J = \{\alpha, \tau \alpha, \tau^2 \alpha,\dots,\tau^k \alpha, \beta\},
    \end{equation*}
    where $\tau^k \alpha < \beta$ and $\tau^{k+1} \alpha \geq \beta$.
\end{definition}
Let $d_\text{min}$ be the length of the shortest edge of $T$ and $d_\text{max}$ be the sum of the lengths of all edges of $T$.
Also, let $J$ be the $(d_\text{min},d_\text{max},1+\varepsilon)$-jump.
Suppose $J = \{\ell_1,\dots,\ell_{|J|}\}$.
Then we do $|J|$ threshold probability queries using the thresholds $\ell_1,\dots,\ell_{|J|}$, and compute
\begin{equation*}
E = \sum_{i=1}^{|J|} C_{\geq \ell_i}(S) \cdot (\ell_i - \ell_{i-1})
\end{equation*}
as an approximation of $\textbf{E}[\kappa(S)]$ (where $\ell_0 = 0$).
Note that $|J| = O(\log_{1+\varepsilon} \frac{d_\text{max}}{d_\text{min}}) = O(\log_{1+\varepsilon} n) = O(\varepsilon^{-1} \log n)$.
It is easy to verify that $E \leq \textbf{E}[\kappa(S)] \leq (1+\varepsilon)E$.

The problem becomes interesting when the spread of $T$ is unbounded.
In this case, although the above method still correctly approximates $\textbf{E}[\kappa(S)]$, the number of the threshold probability queries is no longer well bounded.
Imagine that the $O(n^2)$ possible values of $\kappa(S)$ are distributed as $\ell$, $(1+\varepsilon)\ell$, $(1+\varepsilon)^2\ell$, etc.
Then the $(d_\text{min},d_\text{max},1+\varepsilon)$-jump $J$ is of size $\Omega(n^2)$.
Moreover, for guaranteeing the correctness, it seems that we cannot ``skip'' any element in $J$.
However, as one will realize later, such an extreme situation can never happen.
Recall that we are working on a weighted tree and the $O(n^2)$ possible values of $\kappa(S)$ are indeed the pairwise distances of the vertices of the tree.
As such, these values are not arbitrary, and our insight here is to exploit the underlying properties of the distribution of these values.

Let $e_1,\dots,e_{n-1}$ be the edges of $T$ where $e_i$ has the length (weight) $w_i$.
Assume $w_1 \leq \cdots \leq w_{n-1}$.
We define an index set
\begin{equation*}
I = \left\{m: \sum_{i=1}^{m-1} w_i < w_m \right\}.
\end{equation*}
Suppose $I = \{m_1,\dots,m_k\}$ where $m_1 < \cdots < m_k$.
Note that $m_1 = 1$. 
For convenience, we set $m_{k+1} = n$.
We design our threshold probability queries as follows.
Let $J_i$ be the $(w_{m_i}, s_i, 1+\varepsilon)$-jump where $s_i = \sum_{j<m_{i+1}} w_j$, and $J = J_1 \cup \cdots \cup J_k$.
Suppose $J = \{\ell_1,\dots,\ell_{|J|}\}$ and set $\ell_0 = 0$.
Similarly to the previous case, we do $|J|$ threshold probability queries using the thresholds $\ell_1,\dots,\ell_{|J|}$, and compute
\begin{equation*}
E = \sum_{i=1}^{|J|} C_{\geq \ell_i}(S) \cdot (\ell_i - \ell_{i-1})
\end{equation*}
as an approximation of $\textbf{E}[\kappa(S)]$.
We first verify the correctness, i.e., $E \leq \textbf{E}[\kappa(S)] \leq (1+\varepsilon)E$.
The fact $E \leq \textbf{E}[\kappa(S)]$ can be easily verified.
To see the inequality $\textbf{E}[\kappa(S)] \leq (1+\varepsilon)E$, we define a piecewise-constant function $h:\mathbb{R}^+ \cup \{0\} \rightarrow [0,1]$ as
\begin{equation*}
h(\ell) = \left\{
\begin{array}{ll}
C_{\geq \ell_i}(S) & \text{if } (1+\varepsilon) \ell_i < \ell \leq (1+\varepsilon) \ell_{i+1}, \\
0 & \text{if } \ell>(1+\varepsilon)l_{|J|}, \\
1 & \text{if } \ell=0.
\end{array}
\right.
\end{equation*}
Then it is clear that $(1+\varepsilon) E = \int_{0}^{\infty} h(\ell) d\ell$.
We claim that $\int_{0}^{\infty} h(\ell) d\ell \geq \int_{0}^{\infty} C_{\geq \ell}(S) d\ell$, whence we have $\textbf{E}[\kappa(S)] \leq (1+\varepsilon)E$.
Note that the jumps $J_1,\dots,J_k$ are disjoint and each of them contains a consecutive portion of the sequence $\ell_1,\dots,\ell_{|J|}$.
Furthermore, if $\ell_i$ and $\ell_{i+1}$ belong to different jumps, then there is no possible value of $\kappa(S)$ within the range $(\ell_i,\ell_{i+1})$, i.e., $C_{\geq \ell}(S)$ is constant when $\ell \in [\ell_i,\ell_{i+1})$.
With this observation, it is not difficult to verify that $h(\ell) \geq C_{\geq \ell}(S)$ for any $\ell \geq 0$.
Consequently, we have $\textbf{E}[\kappa(S)] \leq (1+\varepsilon)E$, which implies the correctness of our method.
Now the only thing remaining is to bound the number of the threshold probability queries.
\begin{theorem} \label{numofqueries}
    For each jump $J_i$, we have $|J_i| = O(\varepsilon^{-1}(m_{i+1}-m_i))$.
    As a result, the total number of the threshold probability queries, $|J|$, is $O(\varepsilon^{-1}n)$.
    \textnormal{(See Appendix~\ref{prfnumofqueries} for a proof.)}
\end{theorem}
Indeed, the above method can be extended to a much more general case, in which the stochastic dataset $S$ is given in any metric space $\mathcal{X}$ (not necessarily a tree space).
In this case, one can still define the threshold probability $C_{\geq \ell}(S)$ as well as the expected closest-pair distance $\textbf{E}[\kappa(S)]$ in the same fashion.
Our conclusion is the following.
\begin{theorem} \label{anymetric}
    Given a set $S$ of $n$ stochastic points in a metric space $\mathcal{X}$, one can $(1+\varepsilon)$-approximate the expected closest-pair distance of $S$, $\textnormal{\textbf{E}}[\kappa(S)]$, via $O(\varepsilon^{-1}n)$ threshold probability queries.
    \textnormal{(See Appendix~\ref{prfanymetric} for a proof.)}
\end{theorem}
Concerning the expected closest-pair distance in tree space, we can eventually conclude the following by plugging in our algorithm in Section~\ref{thpr} for computing $C_{\geq \ell}(S)$.
\begin{corollary}
    Given a tree space $\mathcal{T}$ represented by a weighted tree $T$ with $t$ vertices and a set $S$ of $n$ stochastic points in $\mathcal{T}$, one can compute a $(1+\varepsilon)$-approximation for the expected closest-pair distance of $S$, $\textnormal{\textbf{E}}[\kappa(S)]$, in $O(t + \varepsilon^{-1}\min\{tn^2,n^3\})$ time.
\end{corollary}

\section{The most-likely nearest-neighbor search}
In this section, we study the $k$ most-likely nearest-neighbor ($k$-LNN) search in a tree space.
Again, let $\mathcal{T}$ be a tree space represented by a $t$-vertex weighted tree $T$ and $S = \{a_1,\dots,a_n\} \subset \mathcal{T}$ be the given stochastic dataset where the point $a_i$ has an existence probability $\pi_{a_i}$.
The $k$-LNN search problem can be defined as follows.
Let $q \in \mathcal{T}$ be any point.
For each $a_i \in S$, define $\mathit{NNP}_q(a_i)$ as the probability that the nearest-neighbor of $q$ in a realization of $S$ is $a_i$.
Clearly, the nearest-neighbor of $q$ in a realization is $a_i$ iff $a_i$ is in the realzation and any point closer to $q$ is not in the realization.
Therefore, we have
\begin{equation*}
\mathit{NNP}_q(a_i) = \pi_{a_i} \cdot \prod_{x \in \varGamma} (1-\pi_x),
\end{equation*}
where $\varGamma = \{x \in S: \mathit{dist}(q,x) < \mathit{dist}(q,a_i)\}$.
Given a query point $q \in \mathcal{T}$, the goal of the $k$-LNN search is to report the $k$-LNN of $q$, which is a $k$-sequence $(a_{i_1},\dots,a_{i_k})$ of points in $S$ such that $\mathit{NNP}_q(a_{i_1}) \geq \cdots \geq \mathit{NNP}_q(a_{i_k}) \geq \mathit{NNP}_q(a_j)$ for all $j \notin \{i_1,\dots,i_k\}$.
For convenience, we assume $\mathit{NNP}_q(a_i) \neq \mathit{NNP}_q(a_j)$ for any $q \in \mathcal{T}$ and $a_i \neq a_j$ so that the $k$-LNN of any query point $q \in \mathcal{T}$ is uniquely defined.

\begin{figure}[htpb]
    \centering
    \includegraphics[height=4cm]{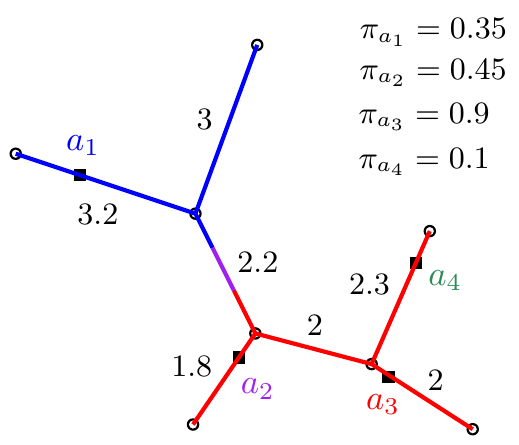}
    \caption{A tree-space 1-LVD with 3 cells.}
    \label{fig:6}
\end{figure}

A standard tool for nearest-neighbor search is the Voronoi diagram.
In stochastic setting, we seek the most-likely Voronoi diagram (LVD), the concept of which is for the first time introduced in \cite{suri2014most}.
The $k$-LVD partitions the query space into connected cells such that points in the same cell have the same $k$-LNN.
Figure~\ref{fig:6} presents an example of $1$-LVD in a tree space.

\subsection{The size of the tree-space LVD} \label{sec_sizeoflvd}
\begin{figure}[htpb]
    \centering
    \includegraphics[height=4cm]{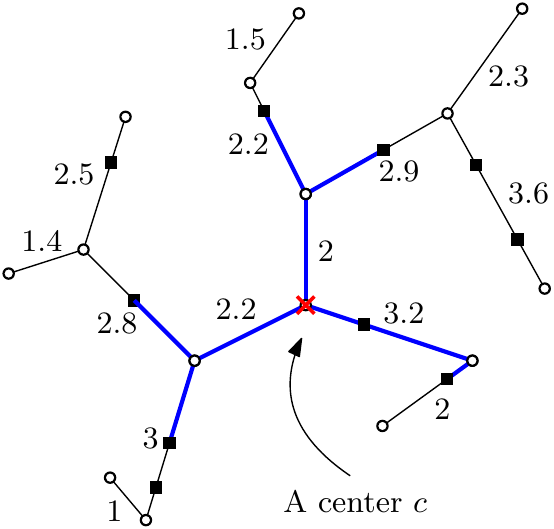}
    \caption{A degree-3 center involving 5 points.}
    \label{fig:4}
\end{figure}

We use $\varPsi_\mathcal{T}^S$ to denote the $k$-LVD of $S$ on $\mathcal{T}$, i.e., the collection of the cells.
Formally, $\varPsi_\mathcal{T}^S$ can be defined as follows.
For any $k$-sequence $\eta = (a_{i_1},\dots,a_{i_k})$, let $\varPsi_\eta$ be the set of the connected components of the subspace $\{q \in \mathcal{T}: \eta \text{ is the } k \text{-LNN of }q\}$.
Then $\varPsi_\mathcal{T}^S$ is the union of $\varPsi_\eta$ over all possible $\eta$.
Clearly, the size of $\varPsi_\mathcal{T}^S$ significantly influences the space efficiency of the LVD-based algorithm for $k$-LNN search.
Let $m_{ij} \in \mathcal{T}$ be the ``midpoint'' of $a_i$ and $a_j$, i.e., the midpoint of the path between $a_i$ and $a_j$ in $\mathcal{T}$.
It is easy to see that the $k$-LNN only changes nearby these $\binom{n}{2}$ midpoints.
However, this does not immediately imply that the size of $\varPsi_\mathcal{T}^S$ is bounded by $O(n^2)$.
The reason is that $O(n^2)$ points do not necessarily decompose $\mathcal{T}$ into $O(n^2)$ pieces (cells), unless these points only locate in the interiors of the edges.
Note that throughout this section, we do not make any spatial assumption about the midpoints.
In other words, it is allowed that different midpoints occupy the same location in $\mathcal{T}$, and some midpoints locate at the vertices of $T$.
The reason why we allow this is explained in Appendix~\ref{remark}.
It is not surprising that even in such a general setting, the size of $\varPsi_\mathcal{T}^S$ is still bounded by $O(n^2)$.
We will see this later as a direct corollary of a technical result (Lemma~\ref{wstsize}).
\begin{definition}
    For any two midpoints $m_{ij}$ and $m_{i'j'}$, we define $m_{ij} \equiv m_{i'j'}$ iff $m_{ij}$ and $m_{i'j'}$ have the same location in $\mathcal{T}$ and $\mathit{dist}(a_i,m_{ij}) = \mathit{dist}(a_j,m_{ij}) = \mathit{dist}(a_{i'},m_{i'j'}) = \mathit{dist}(a_{j'},m_{i'j'})$.
    Clearly, $\equiv$ is an equivalence relation over the midpoints.
    We call the equivalence classes under $\equiv$ \textbf{centers} of $S$ and use $[m_{ij}]$ to denote the center that contains $m_{ij}$.
    A stochastic points $a_i \in S$ is said to be \textbf{involved} by a center $c$ if $c = [m_{ij}]$ for some $j$.
    The \textbf{degree} of a center $c$, denoted by $\mathit{deg}(c)$, is defined as the number of the connected components of $\mathcal{T} \backslash \hat{c}$ that contain at least one point involved by $c$, where $\hat{c}$ denotes the point in $\mathcal{T}$ corresponding to $c$, and each such component is called a \textbf{branch} of $c$.
    A center $c$ is said to be \textbf{critical} if $\hat{c}$ is not in the interior of any cell $C \in \varPsi_\mathcal{T}^S$ and there exists at least one point involved by $c$ that is in the $k$-LNN of $\hat{c}$.
    \textnormal{(See Figure~\ref{fig:4} for an intuitive illustration of center.)}
\end{definition}
\begin{lemma} \label{wstsize}
    Let $\varGamma$ be the set of the critical centers and $\xi = \sum_{c \in \varGamma} \mathit{deg}(c)$.
    Then $|\varPsi_\mathcal{T}^S| \leq \xi+1$.
\end{lemma}
The above lemma immediately gives us the $O(n^2)$ upper bound for the size of $\varPsi_\mathcal{T}^S$.
Indeed, a center $c$ of $S$ contains at least $\Omega(\mathit{deg}(c) \cdot m)$ midpoints, where $m$ is the number of the points involved by $c$, so $\xi+1$ is at most $O(n^2)$.
Unfortunately, this upper bound is tight, following from the $\Omega(n^2)$ worst-case lower bound for the size of the 1-dim 1-LVD given by \cite{suri2014most} (note that the 1-dim LVD is a special case of the tree-space LVD).
Surprisingly, we show that, if we make reasonable assumptions for the existence probabilities of the stochastic points or consider the average case, the size of $\varPsi_\mathcal{T}^S$ is significantly smaller.
Our results are the following.
\begin{itemize}
    \item If the existence probabilities of all points in $S$ are \textit{constant-far from} 0, i.e., there is a fixed constant $\varepsilon>0$ such that $\pi_{a_i} \geq \varepsilon$ for all $a_i \in S$, then the size of the $k$-LVD $\varPsi_\mathcal{T}^S$ is $O(kn)$.
    Note that this assumption about the existence probabilities is natural and reasonable.
    In applications, an extremely small existence probability means the data point is highly unreliable.
    Such a point can be considered as a noise and removed from the dataset.
    \item The average-case size of the $k$-LVD $\varPsi_\mathcal{T}^S$ is $O(kn)$.
    For the average-case analysis we assume that the existence probabilities of the points in $S$ are i.i.d. random variables drawn from any fixed distribution (e.g., the uniform distribution among $[0,1]$).
    In other words, we consider the expectation of $|\varPsi_\mathcal{T}^S|$ when $\pi_{a_1},\dots,\pi_{a_n}$ are such random variables.
    The interesting point is that the $O(kn)$ upper bound is totally independent of the structure of $\mathcal{T}$ and the locations of the stochastic points.
    The randomness is only applied to the existence probabilities in our average-case analysis.
\end{itemize}
To prove these bounds requires new ideas.
By Lemma~\ref{wstsize}, to bound the size of $\varPsi_\mathcal{T}^S$, it suffices to bound the degree-sum of the critical centers.
Intuitively, if a center $c$ is far from the points it involves (compared with other points in $S$), then $c$ is less likely to be critical, as the $c$-involved points are less likely to be in the $k$-LNN of $\hat{c}$.
Along with this intuition, we define the following.
\begin{definition}
    For any center $c$, the \textbf{diameter} of $c$, denoted by $\mathit{diam}(c)$, is defined as the distance from $\hat{c}$ to the $c$-involved points.
    Let $A \subset \mathcal{T}$ be a finite set.
    We define the \textbf{depth} of $c$ with respect to $A$ as $\mathit{dep}_A(c) = |\{x \in A: \mathit{dist}(x,c) < \mathit{diam}(c)\}|$, i.e., the number of the points in $A$ which are closer to $c$ than the $c$-involved points.
\end{definition}
Our idea here is to first bound the ``contribution'' (degree-sum) of the ``shallow'' centers, and then further bound the degree-sum of the critical centers.
Specifically, we investigate the degree-sum of the $d$-\textit{shallow centers} of $S$, i.e., the centers of depth less than $d$ with respect to $S$.
\begin{lemma} \label{numofshallow}
    For $1 \leq d \leq n-1$, the degree-sum of the $d$-shallow centers of $S$ is at most $8dn$.
\end{lemma}
Now we are ready to prove the $O(kn)$ bound for $|\varPsi_\mathcal{T}^S|$ under the ``constant-far from 0'' assumption about the existence probabilities.
\begin{lemma} \label{okshallow}
    If the existence probabilities of the points in $S$ are constant-far from 0, then a center of $S$ is critical only if it is $O(k)$-shallow.
\end{lemma}
\begin{theorem} \label{bound1}
    If the existence probabilities of the points in $S$ are constant-far from 0, then the size of the $k$-LVD $\varPsi_\mathcal{T}^S$ is $O(kn)$.
\end{theorem}
\textit{Proof.}
Suppose the existence probabilities $\pi_{a_1},\dots,\pi_{a_n}$ are constant-far from 0.
Lemma \ref{okshallow} shows that all the critical centers of $S$ are $O(k)$-shallow.
By further applying Lemma~\ref{numofshallow}, the degree-sum of the critical centers is $O(kn)$.
Finally, by Lemma~\ref{wstsize}, the size of $\varPsi_\mathcal{T}^S$ is $O(kn)$.
\hfill $\Box$
\smallskip

\noindent
To prove the bound for the average-case size requires more efforts.
Let $f$ be a fixed probability distribution function whose support is in $(0,1]$ and $\mu$ be the supremum of the support of $f$.
Define $\mu_0 = \mu/(1+\mu)$ and $\lambda = 1-\int_{-\infty}^{\mu_0}f(x) dx$.
For convenience, here we assume $f$ is a continuous distribution (if $f$ is discrete, $\lambda$ can be defined similarly by replacing the integration with a summation).
Clearly, if $X$ is a random variable drawn from $f$, then $\lambda = \Pr[X>\mu_0]$.
Note that $\lambda$ is always positive by definition.
The following lemma clarifies the meaning of $\mu_0$.
\begin{lemma} \label{condforcrit}
    Suppose $\pi_{a_1},\dots,\pi_{a_n}$ are i.i.d. random variables drawn from $f$.
    For any center $c$ of $S$, the event ``$c$ is critical'' does \textbf{not} happen if there are $k$ (distinct) points $a_{i_1},\dots,a_{i_k}$ in $S$ closer to $\hat{c}$ than the $c$-involved points such that $\pi_{a_{i_1}},\dots,\pi_{a_{i_k}}$ are greater than $\mu_0$.
\end{lemma}
\begin{theorem} \label{bound2}
    The average-case size of  $\varPsi_\mathcal{T}^S$ is $O(kn)$, provided that the existence probabilities of the points in $S$ are i.i.d. random variables drawn from a fixed distribution.
\end{theorem}
\textit{Proof.}
Suppose the existence probabilities $\pi_{a_1},\dots,\pi_{a_n}$ are drawn independently from $f$.
Lemma~\ref{condforcrit} implies that, if $c$ is a center of $S$ with $\mathit{dep}_S(c) = d \geq k$, then
\begin{equation*}
\Pr [c \text{ is critical}] \leq u_d = \sum_{i=0}^{k-1} \binom{d}{i} \lambda^i (1-\lambda)^{d-i}. 
\end{equation*}
Then by applying Lemma~\ref{wstsize}, we have
\begin{equation*}
\mathbf{E}[|\varPsi_\mathcal{T}^S|] \leq \sum_c \Pr [c \text{ is critical}] \cdot \mathit{deg}(c) \leq \sum_{c \in H_k} \mathit{deg}(c) + \sum_{d=k+1}^{n-1} \sum_{c \in H_d} (u_{d-1} - u_d) \mathit{deg}(c),
\end{equation*}
where $H_d$ is the set of the $d$-shallow centers of $S$.
Observe that
\begin{equation*}
u_{d-1} - u_d = \binom{d-1}{k-1} \lambda^k (1-\lambda)^{d-k}.
\end{equation*}
Based on this and Lemma~\ref{numofshallow}, we further have
\begin{equation*}
\mathbf{E}[|\varPsi_\mathcal{T}^S|] \leq 8kn + 8n \sum_{d=k+1}^{n-1} \binom{d-1}{k-1} \lambda^k (1-\lambda)^{d-k} d.
\end{equation*}
Note that
\begin{equation*}
\sum_{d=k+1}^{n-1} \binom{d-1}{k-1} \lambda^k (1-\lambda)^{d-k} d = k \left( \frac{\lambda}{1- \lambda}\right)^k \sum_{d=k+1}^{n-1} \binom{d}{k} (1-\lambda)^d.
\end{equation*}
By an induction argument on $k$, it is not difficult to see that
\begin{equation*}
\sum_{d=k+1}^{n-1} \binom{d}{k} (1-\lambda)^d < \sum_{d=k}^\infty \binom{d}{k} (1-\lambda)^d = \frac{(1-\lambda)^{k}}{\lambda^{k+1}}.
\end{equation*}
Finally, by combining the inequalities, we have $\mathbf{E}[|\varPsi_\mathcal{T}^S|] \leq 8kn + \frac{8kn}{\lambda} = O(kn)$.
\hfill $\Box$

\subsection{Constructing LVD and answering queries}
In this section, we show how to construct the $k$-LVD $\varPsi_\mathcal{T}^S$ and use it to answer $k$-LNN queries.
Let $e_1,\dots,e_{t-1}$ be the edges of $T$.
Assume each edge $e_i$ has a specified ``start point'' $s_i$ (which is one of its two endpoints) and the query point $q$ is specified via a pair $(i,\delta)$ meaning the point on $e_i$ with distance $\delta$ to $s_i$.

We first explain the data structure used for storing the $k$-LVD $\varPsi_\mathcal{T}^S$ and answering queries.
The LVD data structure is simple.
First, it contains $|\varPsi_\mathcal{T}^S|$ arrays (called \textit{answer arrays}) each of which stores the $k$-LNN answer of one cell of $\varPsi_\mathcal{T}^S$.
This part takes $O(k|\varPsi_\mathcal{T}^S|)$ space.
In addition to the answer arrays, we also need to record the structure of $\varPsi_\mathcal{T}^S$.
For each edge $e_i$ of $T$, we use a sorted list $L_i$ to store the ``cell-decomposition'' of $e_i$, i.e., how $\varPsi_\mathcal{T}^S$ decomposes $e_i$.
Specifically, the intersection of each cell $C \in \varPsi_\mathcal{T}^S$ and $e_i$ is an ``interval'' (may be empty).
These intervals (associated with the corresponding cell-labels) are stored in $L_i$ in the order they appear on $e_i$.
Note that this part takes $O(t+|\varPsi_\mathcal{T}^S|)$ space.
Indeed, if an edge is decomposed into $p$ pieces (intervals) by $\varPsi_\mathcal{T}^S$, then it at least entirely contains $(p-2)$ cells of $\varPsi_\mathcal{T}^S$ (so we can charge these $(p-2)$ pieces to the corresponding cells and the remaining two pieces to the edge).
Therefore, the total space of the LVD data structure is $O(t+k|\varPsi_\mathcal{T}^S|)$.
To answer a query $q = (i,\delta)$, we first do a binary search in the list $L_i$ to know which cell $q$ locates in.
Then we use the answer array corresponding to the cell to output the $k$-LNN of $q$ directly.
The time cost for answering the query is clearly $O(\log |\varPsi_\mathcal{T}^S| + k)$.

\begin{figure}[htpb]
    \centering
    \includegraphics[height=4cm]{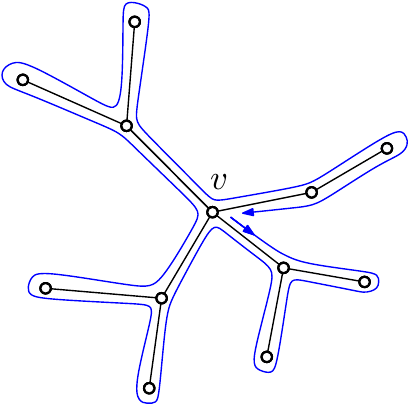}
    \caption{A walk in tree visiting each edge exactly twice.}
    \label{fig:5}
\end{figure}

Next, we consider the construction of the LVD data structure.
Due to limited space, we only present the main procedure of the construction algorithm, and defer the details to Appendix~\ref{detconstr}.
The first step of the construction is to compute all the centers of $S$ and sort the centers in the interior of each edge $e$ in the order they appear on $e$.
We are able to get this done in $O(t+n^2 \log n)$ time (see Appendix~\ref{compmid}).
After the centers are computed and sorted, we begin to construct the LVD data structure.
Choose a vertex $v$ of $T$.
Starting at $v$, we do a walk in $\mathcal{T}$ along with the edges of $T$.
The walk visits each edge of $T$ exactly twice and finally goes back to $v$.
See Figure~\ref{fig:5} for an illustration of the walk.
During the walk, we maintain a (balanced) binary search tree for the nearest-neighbor probabilities of $a_1,\dots,a_n$ with respect to the current location $x$, i.e., $\mathit{NNP}_x(a_1),\dots,\mathit{NNP}_x(a_n)$.
By exploiting this binary search tree, we can work out the cell-decomposition of each edge $e_i$ (i.e., the sorted list $L_i$) at the first time we visit $e_i$ in the walk.
Specifically, we track the $k$-LNN when walking along with $e_i$, which can be obtained by retrieving the $k$ largest elements from the binary search tree.
Whenever the $k$-LNN changes, a new cell of $\varPsi_\mathcal{T}^S$ is found, so we need to create a new answer array to store the $k$-LNN information.
Also, we need to update the sorted list $L_i$.
In this way, after we go through $e_i$ (for the first time), the computation of $L_i$ is completed.
At the second time we visit an edge, we do nothing but maintain the binary search tree.
When we finish the walk and go back to $v$, the construction of the LVD data structure is done.
Clearly, in the process of the walk, we only need to maintain the binary search tree and retrieve the $k$-LNN when we arrive at (resp., leave from) a center of $S$ from (resp., to) one of its branches.
With a careful implementation and analysis, we can complete the work for each center $c$ in $O(\mathit{deg}(c) \cdot m_c \log n +\mathit{deg}(c) \cdot k)$ time, where $m_c$ is the number of the stochastic points involved by $c$ (see Appendix~\ref{walk} for details).
Thus, the total time cost for maintaining binary search tree and retrieving $k$-LNN is $O(n^2 \log n + n^2 k)$.
It follows that the entire walk can be completed in $O(t+ n^2 \log n + n^2 k)$ time, which is also the overall construction time for the LVD data structure.
Combined with the bounds for the size of the tree-space LVD proved in Section~\ref{sec_sizeoflvd}, we then have the following results.
\begin{theorem}
    Given a tree space $\mathcal{T}$ represented by a $t$-vertex weighted tree and a set $S$ of $n$ stochastic points in $\mathcal{T}$, one can construct in $O(t+ n^2 \log n + n^2 k)$ time an LVD data structure to answer $k$-LNN queries in $O(\log n + k)$ time.
    The LVD data structure uses worst-case $O(t+kn^2)$ space and average-case $O(t+k^2n)$ space.
    Furthermore, if the existence probabilities of the points in $S$ are constant-far from 0, then the LVD data structure uses worst-case $O(t+k^2n)$ space.
\end{theorem}

\bibliography{my_bib}

\begin{thebibliography}{10}

\bibitem{agarwal2013nearest}
P.K. Agarwal, B.~Aronov, S.~Har-Peled, J.M. Phillips, K.~Yi, and W.~Zhang.
\newblock Nearest neighbor searching under uncertainty {II}.
\newblock In {\em Proc. of the 32nd Sympos. on PODS}, pages 115--126. ACM,
  2013.

\bibitem{agarwal2012range}
P.K. Agarwal, S-W. Cheng, and K.~Yi.
\newblock Range searching on uncertain data.
\newblock {\em ACM Transactions on Algorithms}, 8(4):43, 2012.

\bibitem{agarwal2012nearest}
P.K. Agarwal, A.~Efrat, S.~Sankararaman, and W.~Zhang.
\newblock Nearest-neighbor searching under uncertainty.
\newblock In {\em 31st SIGMOD/PODS}. ACM, 2012.

\bibitem{agarwal2014convex}
P.K. Agarwal, S.~Har-Peled, S.~Suri, H.~Y{\i}ld{\i}z, and W.~Zhang.
\newblock Convex hulls under uncertainty.
\newblock In {\em Algorithms-ESA}, pages 37--48. Springer, 2014.

\bibitem{agarwal2016range}
P.K. Agarwal, N.~Kumar, S.~Sintos, and S.~Suri.
\newblock Range-max queries on uncertain data.
\newblock In {\em Proc. of the 35th SIGMOD/PODS}, pages 465--476. ACM, 2016.

\bibitem{chen2013efficient}
J.~Chen and L.~Feng.
\newblock Efficient pruning algorithm for top-k ranking on dataset with value
  uncertainty.
\newblock In {\em Proc. of the 22nd CIKM}, pages 2231--2236. ACM, 2013.

\bibitem{fink2016hyperplane}
M.~Fink, J.~Hershberger, N.~Kumar, and S.~Suri.
\newblock Hyperplane separability and convexity of probabilistic point sets.
\newblock In {\em Proc. of the 32nd SoCG}. ACM, 2016.

\bibitem{ge2009top}
T.~Ge, S.~Zdonik, and S.~Madden.
\newblock Top-$k$ queries on uncertain data: on score distribution and typical
  answers.
\newblock In {\em Proc. of the 2009 SIGMOD}, pages 375--388. ACM, 2009.

\bibitem{huang2015approximating}
L.~Huang and J.~Li.
\newblock Approximating the expected values for combinatorial optimization
  problems over stochastic points.
\newblock In {\em Intl. Colloquium on Automata, Languages, and Programming},
  pages 910--921. Springer, 2015.

\bibitem{kamousi2011stochastic2}
P.~Kamousi, T.M. Chan, and S.~Suri.
\newblock Stochastic minimum spanning trees in euclidean spaces.
\newblock In {\em Proc. of the 27th SoCG}, pages 65--74. ACM, 2011.

\bibitem{kamousi2014closest}
P.~Kamousi, T.M. Chan, and S.~Suri.
\newblock Closest pair and the post office problem for stochastic points.
\newblock {\em Computational Geometry}, 47(2):214--223, 2014.

\bibitem{loffler2010largest}
M.~L{\"o}ffler and M.~van Kreveld.
\newblock Largest and smallest convex hulls for imprecise points.
\newblock {\em Algorithmica}, 56(2):235--269, 2010.

\bibitem{reem2011geometric}
D.~Reem.
\newblock The geometric stability of voronoi diagrams with respect to small
  changes of the sites.
\newblock In {\em Proc. of the 27th SoCG}, pages 254--263. ACM, 2011.

\bibitem{suri2014most}
S.~Suri and K.~Verbeek.
\newblock On the most likely {V}oronoi {D}iagramand nearest neighbor searching.
\newblock In {\em ISAAC}, pages 338--350. Springer, 2014.

\bibitem{suri2013most}
S.~Suri, K.~Verbeek, and H.~Y{\i}ld{\i}z.
\newblock On the most likely convex hull of uncertain points.
\newblock In {\em Algorithms--ESA}, pages 791--802. Springer, 2013.

\bibitem{xue2016separability}
J.~Xue, Y.~Li, and R.~Janardan.
\newblock On the separability of stochastic geometric objects, with
  applications.
\newblock In {\em Proc. of the 32nd SoCG}. ACM, 2016.

\end{thebibliography}

\newpage\appendix{\noindent\bf\huge Appendix}

\section{Missing proofs} \label{prfs}
\subsection{Proof of Theorem~\ref{preproc}} \label{prfpreproc}
Clearly, we can represent $\mathcal{T}$ by a new tree $T'$ with $O(t+n)$ vertices such that each stochastic point in $S$ lies at a vertex of $T'$.
The tree $T'$ is obtained by adding some new vertices to $T$ for the stochastic points lying in the interiors of the edges and ``breaking'' those edges.
It can be easily computed in $O(t+n \log n)$ time by sorting the stochastic points in the interior of each edge (in the order they appear on the edge).
Next, we try to simplify $T'$ to make it have $O(n)$ vertices.
We say a vertex of $T'$ is \textit{empty} if there is no stochastic point lying at it.
The first step is to delete the branches of $T'$ which do not contain any stochastic points.
Specifically, if $T'$ has an empty leaf $v$, we then remove $v$ and its adjacent edge from $T'$.
Keep doing this until $T'$ has no empty leaf.
After this step, the underlying tree space of $T'$ changes to be a subspace of the original $\mathcal{T}$.
The second step is to compress the ``empty chains'' in $T'$.
Specifically, if $T'$ has a degree-2 empty vertex $v$ with edges $e_1 = (v,v')$ and $e_2 = (v,v'')$, we replace $v, e_1, e_2$ with a single edge $e = (v',v'')$ whose weight is the sum of the weights of $e_1$ and $e_2$.
Note that this operation does not change the underlying tree space.
We keep doing this until $T'$ has no degree-2 empty vertex.
These two steps of simplification can be done in $O(t+n)$ time. 
In the resulting $T'$, every empty vertex has a degree at least 3.
Thus, $T'$ has $O(n)$ vertices.
Furthermore, $T'$ represents a tree space $\mathcal{T}'$ such that $S \subset \mathcal{T}' \subseteq \mathcal{T}$ and each stochastic point in $S$ is located at a vertex of $T'$.
\subsection{Proof of Lemma~\ref{th1}}
The ``only if'' part is easy to see.
Assume that $S'$ is legal.
Let $x \in S\backslash\{a_1\}$ be any point.
If $x$ does not satisfy the condition~(1) and (2), i.e., $\omega(x,S')$ is defined and $\omega(x,S') \neq \omega(\bar{p}(a_i),S')$, then it must satisfy the condition~(3) because both $\omega(x,S')$ and $\omega(\bar{p}(a_i),S')$ are in $S'$.
To show the ``if'' part, assume that $S'$ is not legal.
Then we can find distinct points $x,y \in S'$ such that $\mathit{dist}(x,y) < \ell$.
Let $z$ be the lowest common ancestor of $x$ and $y$ in $T$.
Without loss of generality, we can assume $x \neq z$.
Suppose $\hat{x}$ is the child of $z$ such that $x \in V(T_{\hat{x}})$.
We consider two cases, $\omega(z,S') \notin V(T_{\hat{x}})$ and $\omega(z,S') \in V(T_{\hat{x}})$ (note that $\omega(z,S')$ is defined since both $x$ and $y$ are in $V(T_z) \cap S'$).
In the case of $\omega(z,S') \notin V(T_{\hat{x}})$, we show that $\hat{x}$ satisfies none of the three conditions.
First, because $x \in V(T_{\hat{x}}) \cap S'$, $\omega(\hat{x},S')$ is clearly defined so that $\hat{x}$ violates the condition~(1).
Second, we have $\omega(\hat{x},S') \neq \omega(\bar{p}(\hat{x}),S')$ since $\omega(\bar{p}(\hat{x}),S') = \omega(z,S') \notin V(T_{\hat{x}})$, which implies that $\hat{x}$ violates the condition~(2).
Thirdly, since $\omega(z,S') \notin V(T_{\hat{x}})$, we have
\begin{equation*}
\mathit{dist}(\omega(\hat{x},S'),\omega(\bar{p}(\hat{x}),S')) = \mathit{dist}(\omega(\hat{x},S'),z) + \mathit{dist}(z,\omega(z,S')). 
\end{equation*}
Furthermore, by the definition of witness, $\mathit{dep}(\omega(\hat{x},S')) \leq \mathit{dep}(x)$ and thus $\mathit{dist}(\omega(\hat{x},S'),z) \leq \mathit{dist}(x,z)$.
Similarly, $\mathit{dep}(\omega(z,S')) \leq \mathit{dep}(y)$ and thus $\mathit{dist}(\bar{p}(\hat{x}),\omega(z,S')) = \mathit{dist}(z,\omega(z,S')) \leq \mathit{dist}(z,y)$.
Note that $\mathit{dist}(x,z)+\mathit{dist}(z,y) = \mathit{dist}(x,y) < l$.
Therefore, we can conclude that $\mathit{dist}(\omega(\hat{x},S'),\omega(\bar{p}(\hat{x}),S')) < l$, which implies that $\hat{x}$ violates the condition (3).
In the case of $\omega(z,S') \in V(T_{\hat{x}})$, we notice that $y \neq z$; otherwise $\omega(z,S')=z \notin V(T_{\hat{x}})$.
Suppose $\hat{y}$ is the child of $z$ such that $y \in V(T_{\hat{y}})$.
Then it is easy to see that $\hat{y}$ satisfies none of the three conditions, by applying the same argument used in the previous case (note that the situation here is dual to the previous case).
\subsection{Proof of Theorem~\ref{comp2}} \label{prfcomp2}
When $y \in V(T_{\bar{p}(x)}) \backslash V(T_x)$, $P_y(x)$ is the probability that a realization $S' \subseteq_\text{R} V(T_x)$ satisfies the conditions that $S' \cup \{y\}$ is legal and $\omega(\bar{p}(x),S' \cup \{y\})=y$.
Clearly, the empty sample $S' = \emptyset$ satisfies the two conditions and its probability is computed by the first term of the formula.
If $S'$ is not empty, then $\omega(x,S')$ is defined and must be some vertex $z \in V(T_x)$.
In this case, we need $y \prec z$ to guarantee $\omega(\bar{p}(x),S' \cup \{y\})=y$.
Also, we need $\mathit{dist}(z,y) \geq \ell$ to ensure the legality of $S' \cup \{y\}$.
Therefore, $z$ must be a vertex in $\varGamma$.
Now it suffices to show that the right-hand side of the formula does not overestimate the probability.
In other words, we want that, if $S' \subseteq V(T_x)$ is legal and $\omega(x,S') = z$ for some $z \in \varGamma$, then $\omega(\bar{p}(x),S' \cup \{y\})=y$ and $S' \cup \{y\}$ is also legal.
The former can be easily seen from the facts that $\omega(x,S') = z$ and $y \prec z$.
To see the latter, by Lemma~\ref{th1}, we only need to verify that $S' \cup \{y\}$ is locally legal at $x$ (the local legalities of $S' \cup \{y\}$ at any vertex other than $x$ is clear).
Note that $z \in \varGamma$, so we have $\mathit{dist}(\omega(x,S'),y) = \mathit{dist}(z,y) \geq \ell$, which completes the proof.
\subsection{Proof of Theorem~\ref{th4}} \label{prfth4}
Clearly, if $y_i \prec z_j$, then $y_i \prec z_{j'}$ for any $j' > j$.
Also, if $\mathit{dist}(z_j,y_i) \geq \ell$, then $\mathit{dist}(z_{j'},y_i) \geq \ell$ for any $j' > j$, because both the paths $z_j \rightarrow y_i$ and $z_{j'} \rightarrow y_i$ go through the vertex $\bar{p}(x)$.
Thus, we know that $\varGamma_y = \{z_j,z_{j+1},\dots,z_m\}$ for some $j \in \{1,\dots,m\}$.
To show the remaining part of the theorem, we notice that $\varGamma_{y_i} = \varGamma'_{y_i} \cap \varGamma''_{y_i}$, where $\varGamma'_{y_i} = \{z \in V(T_x): y_i \prec z \}$ and $\varGamma''_{y_i} = \{z \in V(T_x): \mathit{dist}(z,y_i) \geq \ell \}$.
Both $\varGamma'_{y_i}$ and $\varGamma''_{y_i}$ are suffixes of the sequence $(z_1,\dots,z_m)$.
Furthermore, we have $\varGamma'_{y_1} \supseteq \cdots \supseteq \varGamma'_{y_r}$ and $\varGamma''_{y_1} \subseteq \cdots \subseteq \varGamma''_{y_r}$.
As such, we can conclude that $\varGamma_{y_1} \subseteq \cdots \subseteq \varGamma_{y_k} \supseteq \cdots \supseteq \varGamma_{y_r}$ for some $k \in \{1,\dots,t\}$.
\subsection{Proof of Theorem~\ref{numofnonch}} \label{prfnumofnonch}
Suppose the tree space $\mathcal{T}$ is represented by a $t$-vertex weighted tree $T_0$.
Let $e$ be an edge of $T_0$, and $\hat{e} \subseteq \mathcal{T}$ be the subspace corresponding to $e$.
Assume that $v_1,\dots,v_k$ are the vertices of $T$ lying in $\hat{e}$ (sorted in the order they appear on $\hat{e}$).
We claim that among $v_1,\dots,v_k$, there are only constant number of non-chain vertices.
If the root of $T$ is not in $\{v_1,\dots,v_k\}$, then only $v_1,v_2,v_{k-1},v_k$ can be non-chain vertices.
Otherwise, if the root is some $v_i$, then only $v_1,v_2,v_{i-1},v_i,v_{i+1},v_{k-1},v_k$ can be non-chain vertices.
In both the cases, the number of the non-chain vertices is constant.
Finally, since $T_0$ has $(t-1)$ edges, the total number of the non-chain vertices of $T$ is bounded by $O(t)$.
\subsection{Proof of Theorem~\ref{numofqueries}} \label{prfnumofqueries}
First, for any index $r \in [m_i,m_{i+1})$, we show that $w_r \leq 2^{r-m_i} \cdot w_{m_i}$.
When $r = m_i$, the inequality clearly holds.
Assume for induction that the inequality holds for any index less than $r'$ ($m_i < r' < m_{i+1}$).
Since $r' \notin I$ and $m_i \in I$, we then have
\begin{equation*}
w_{r'} \leq \sum_{j=1}^{r'-1} w_j < w_{m_i} + \sum_{j = m_i}^{r'-1} 2^{j-m_i} \cdot w_{m_i} = 2^{r'-m_i} \cdot w_{m_i},
\end{equation*}
which completes the induction.
It follows that
\begin{equation*}
s_i = \sum_{j<m_{i+1}} w_j < w_{m_i} + \sum_{j = m_i}^{m_{i+1}-1} 2^{j-m_i} \cdot w_{m_i} = 2^{m_{i+1}-m_i} \cdot w_{m_i}.
\end{equation*}
Thus, $|J_i| = O(\log_{1+\varepsilon} \frac{s_i}{w_{m_i}}) = O(\varepsilon^{-1}(m_{i+1}-m_i))$.
Since $|J| = \sum_{i=1}^{k} |J_i|$, we can immediately conclude that $|J| = O(\varepsilon^{-1}n)$.
\subsection{Proof of Theorem~\ref{anymetric}} \label{prfanymetric}
Suppose the stochastic dataset $S = \{a_1,\dots,a_n\}$ is given in a metric space $\mathcal{X}$ with the metric $d_\mathcal{X}$.
Let $G_\mathcal{X}$ be the metric graph of $S$, i.e., a weighted complete graph with vertex-set $S$ such that the weight of each edge $(a_i,a_j)$ is equal to $d_\mathcal{X}(a_i,a_j)$.
Also, let $T$ be a minimum spanning tree of $G_\mathcal{X}$.
We then directly apply the method in Section~\ref{ecdist} to the tree $T$ to compute the quantity $E$ via $O(\varepsilon^{-1}n)$ threshold probability queries.
(Note that the threshold probability queries are made with respect to the metric of $\mathcal{X}$, the tree $T$ is only used for choosing thresholds.)
We show that $E$ gives us a $(1+\varepsilon)$-approximation for $\textbf{E}[\kappa(S)]$.
The fact $E \leq \textbf{E}[\kappa(S)]$ can be easily verified.
To see the inequality $\textbf{E}[\kappa(S)] \leq (1+\varepsilon)E$, we review the analysis in Section~\ref{ecdist}.
Again, we use $e_1,\dots,e_{n-1}$ to denote the edges of $T$ with lengths (weights) $w_1 \leq \cdots \leq w_{n-1}$.
As that in Section~\ref{ecdist}, we have the index set $I = \{m_1,\dots,m_k\}$, the jumps $J_1,\dots,J_k$, and $J = J_1 \cup \cdots \cup J_k = \{\ell_1,\dots,\ell_{|J|}\}$.
Now we only need to verify that if $\ell_i$ and $\ell_{i+1}$ belong to different jumps, then there is no possible value of $\kappa(S)]$ within the range $(\ell_i,\ell_{i+1})$.
As long as this is true, we can use the totally same argument as that in Section~\ref{ecdist} to show $\textbf{E}[\kappa(S)] \leq (1+\varepsilon)E$.
Let $d_T(a_i,a_j)$ be the distance between $a_i$ and $a_j$ in $T$ (i.e., the length of simple simple path between $a_i$ and $a_j$ in $T$).
Assume for contradiction that $\ell_i \in J_r$, $\ell_{i+1} \in J_{r+1}$, and there exists $x,y \in S$ such that $\ell_i < d_\mathcal{X}(x,y) < \ell_{i+1}$.
Observe that $\ell_i = s_r = \sum_{j<m_{r+1}} w_j$ and $\ell_{i+1} = w_{m_{r+1}}$.
Since $d_T(x,y) \geq d_\mathcal{X}(x,y) > \ell_i$, there must be an edge $e_m$ with $m \geq m_{r+1}$ on the path between $x$ and $y$ in $T$.
However, this contradicts the fact that $T$ is a minimum spanning tree, because $d_\mathcal{X}(x,y) < \ell_{i+1}$.
As such, there is no possible value of $\kappa(S)$ within the range $(\ell_i,\ell_{i+1})$.
By applying the analysis in Section~\ref{ecdist}, it turns out that $\textbf{E}[\kappa(S)] \leq (1+\varepsilon)E$.
\subsection{Proof of Lemma~\ref{wstsize}}
Let $x \in \mathcal{T}$ be any point.
We use $B_x$ to denote the (open) $\delta$-ball about $x$ with $\delta$ small enough such that $\hat{c} \in B_x$ only if $\hat{c} = x$ for any center $c$ (not necessarily critical).
We first notice that $\mathit{NNP}_q(a_i) \leq \mathit{NNP}_x(a_i)$ for any $q \in B_x$ and any $a_i \in S$.
This is because if $\mathit{dist}(x,a_j) < \mathit{dist}(x,a_i)$ then $\mathit{dist}(q,a_j) < \mathit{dist}(q,a_i)$.
We further claim that $\mathit{NNP}_q(a_i) < \mathit{NNP}_x(a_i)$ for $q \in B_x$ iff there is a center $c$ (not necessarily critical) with $\hat{c} = x$ such that $a_i$ is involved by $c$ and $q$ is in a branch of $c$ other than the one that contains $a_i$.
To see this, consider a point $a_j \in S$ with $\mathit{dist}(x,a_j) = \mathit{dist}(x,a_i)$ and $\mathit{dist}(q,a_j) < \mathit{dist}(q,a_i)$. Note that such a point always exists, otherwise $\mathit{NNP}_q(a_i) = \mathit{NNP}_x(a_i)$.
It is evident that $q$ and $a_j$ locate in the same connected component of $\mathcal{T} \backslash x$, which is other than the component contains $a_i$.
Thus, the center $c = [m_{ij}]$ satisfies the desired properties.
Now let us prove the theorem.
Recall that $\varGamma$ is the set of the critical centers of $S$.
We show that any connected subspace $U \subseteq \mathcal{T}$ intersecting with (exactly) $p$ cells in $\varPsi_\mathcal{T}^S$ satisfies the condition that $p \leq \sum_{c \in \varGamma, \hat{c} \in U} \mathit{deg}(c)+1$.
When $p=1$, this is trivially true.
Assume that for any $p < p'$ the argument holds, and consider the case $p = p'$.
Let $C$ be a cell satisfying $C \cap U \cap \overline{U \backslash C} \neq \emptyset$.
Note that such a cell always exists, unless $U$ only intersects with one cell and then $p=1$ (as $U$ is connected).
Choose a point $x \in C \cap U \cap \overline{U \backslash C}$ and define $X = \{c \in \varGamma: \hat{c} = x\}$.
Suppose $U \backslash x$ has $l$ connected components $U_1,\dots,U_l$ among which there are $l'$ components not intersecting with $C$.
We denote by $p_i$ the number of the cells in $\varPsi_\mathcal{T}^S$ intersecting with $U_i$.
Then we have
\begin{equation*}
p \leq \sum_{i=1}^{l} p_i - (l - l') + 1.
\end{equation*}
This is because the sum of all $p_i$ counts the cell $C$ exactly $(l-l')$ times and other cells intersecting with $U$ exactly once.
It is easy to observe that $p_i < p$.
Then by our induction hypothesis, we have
\begin{equation*}
\sum_{i=1}^{l} p_i \leq \sum_{c \in \varGamma, \hat{c} \in U \backslash x} \mathit{deg}(c) + l = \sum_{c \in \varGamma \backslash X, \hat{c} \in U} \mathit{deg}(c) + l.
\end{equation*}
Thus, it follows that
\begin{equation*}
p \leq \sum_{c \in \varGamma \backslash X, \hat{c} \in U} \mathit{deg}(c) + l' + 1.
\end{equation*}
It now suffices to show $l' \leq \sum_{c \in X} \mathit{deg}(c)$.
Let $U_i$ be a component not intersecting with $C$ and $q \in U_i \cap B_x$ be any point.
Since $q \notin C$ and $q \in B_x$, $x$ and $q$ have different $k$-LNNs.
As such, there exists a stochastic point $a_j$ in the $k$-LNN of $x$ such that $\mathit{NNP}_q(a_j) < \mathit{NNP}_x(a_j)$ (otherwise $x$ and $q$ have the same $k$-LNN, according to our observation $\mathit{NNP}_q(\cdot) \leq \mathit{NNP}_x(\cdot)$ presented in the beginning of the proof).
Since $\mathit{NNP}_q(a_j) < \mathit{NNP}_x(a_j)$, there is a center $c$ with $\hat{c} = x$ such that $a_j$ is involved by $c$ and $q$ is in one branch of $c$ (again, this follows from our observation in the beginning).
Note that $c \in X$ as it is critical ($c$ involves $a_j$ and $a_j$ is in the $k$-LNN of $x$).
We then charge $U_i$ to the branch of $c$ containing $q$.
We do this for all the $l'$ components not intersecting with $C$.
It is easy to verify that each branch of each center $c \in X$ is charged at most once, which immediately implies that $l' \leq \sum_{c \in X} \mathit{deg}(c)$.
Consequently, the argument holds for $p = p'$ and hence for any $p$.
By setting $U = \mathcal{T}$, we conclude that $|\varPsi_\mathcal{T}^S| \leq \xi+1$.
\subsection{Proof of Lemma~\ref{numofshallow}}
We first prove the special case when $d=1$.
We show that the degree-sum of all the 1-shallow centers (i.e., the centers of depth 0 with respect to $S$) is at most $2n-2$.
If $n=1$, this claim is clearly true, as there is no center.
Assume the claim holds for any $n<n_0$, and consider the case that $n=n_0$.
Let $c$ be a center with $\mathit{dep}_S(c) = 0$.
Suppose $\mathit{deg}(c) = g$ and $S_c \subseteq S$ is the set of points involved by $c$.
Without loss of generality, assume $a_1 \in S_c$.
We observe the following three facts. \\
$\bullet$ For $a_i, a_j \notin S_c$, $\mathit{dep}_S([m_{ij}]) = 0$ only if $a_i$ and $a_j$ are in the same connected components of $\mathcal{T} \backslash \hat{c}$.
To see this, assume that $a_i$ and $a_j$ locate in different connected components.
Then $\mathit{dist}(a_1,m_{ij}) < \mathit{dist}(a_i,m_{ij}) = \mathit{dist}(a_j,m_{ij})$ and hence $\mathit{dep}_S([m_{ij}]) > 0$. \\
$\bullet$ For $a_i \in S_c$ and $a_j \notin S_c$, $\mathit{dep}_S([m_{ij}]) = 0$ only if $a_i$ and $a_j$ are in the same connected component of $\mathcal{T} \backslash \hat{c}$, or $a_j$ is not in any branch of $c$.
To see this, assume $a_i$ and $a_j$ are located in different connected components of $\mathcal{T} \backslash \hat{c}$ and $a_j$ is in the branch of $c$ containing $a_1$ (without loss of generality).
Then $\mathit{dist}(a_1,m_{ij}) < \mathit{dist}(a_i,m_{ij}) = \mathit{dist}(a_j,m_{ij})$ and hence $\mathit{dep}_S([m_{ij}]) > 0$. \\
$\bullet$ Let $a_i \notin S_c$ be a point which does not locate in any branch of $c$.
Then the degree of the center $[m_{1i}]$ does not change if we ``delete'' all the points in $S_c \backslash \{a_1\}$.
Formally, set $S' = S \backslash S_c \cup \{a_1\}$ and denote by $[m_{1i}']$ the center of $S'$ that contains the midpoint of $a_1$ and $a_i$.
Then $\mathit{deg}([m_{ij}]) = \mathit{deg}([m_{ij}'])$.
This observation follows immediately from the fact that all the points in $S_c$ locate in the same connected components of $\mathcal{T} \backslash m_{1i}$. \\
With these observations, we now bound the degree-sum of the 1-shallow centers of $S$ (denoted by $\phi$).
Suppose that $\mathcal{T} \backslash \hat{c}$ has $p$ connected components $U_1,\dots,U_p$, where $S \cap U_i = R_i$.
If $U_i$ is a branch of $c$, we use $\lambda_i$ to denote the degree-sum of the 1-shallow centers of $R_i$, otherwise $\lambda_i$ denotes the degree-sum of the 1-shallow centers of $R_i \cup \{a_1\}$ (here the depths of the considered centers are with respect to $R_i$ or $R_i \cup \{a_1\}$ instead of $S$).
Based on the above three observations and the induction hypothesis, we have
\begin{equation*}
\phi \leq \sum_{i=1}^{p} \lambda_i + g \leq 2 \sum_{i=1}^{p} |R_i|-2g + g \leq 2n-g \leq 2n-2.
\end{equation*}
Thus, the case of $d=1$ is verified.
To prove the result for a general $d$, we use the sampling argument.
We sample each point in $S$ independently with probability $1/d$.
Let $S'$ be the resulting random sample and $\varphi$ be a random variable indicating the degree-sum of the 1-shallow centers of $S'$ (the depths of the considered centers are with respect to $S'$).
The previous proof for $d=1$ implies that $\mathbf{E}[\varphi] \leq 2n/d$.
Clearly, each center of $S'$ is ``contributed'' by some center of $S$.
For each center $c$ of $S$, define a random variable $\sigma(c)$ such that $\sigma(c) = 0$ if $c$ does not contribute a 1-shallow center of $S'$, and $\sigma(c) = \mathit{deg}(c')$ if $c$ contributes a 1-shallow center $c'$ of $S'$.
The event $\sigma(c) = 0$ happens whenever there are at most one point involved by $c$ being sampled to $S'$, or there are points closer to $\hat{c}$ (than those involved by $c$) being sampled to $S'$.
We claim that, for any $d$-shallow center $c$ of $S$, $\mathbf{E}[\sigma(c)] = \Omega (\mathit{deg}(c)/d^2)$.
To see this, we set $g = \mathit{deg}(c)$ and $\theta = \mathit{dep}_S(c) < d$.
Without loss of generality, assume $a_1,\dots,a_g \in S$ are involved by $c$ and belong to distinct branches of $c$.
Define another random variable $\tau$ such that $\tau = |S' \cap \{a_1,\dots,a_g\}|$ if $c$ contributes a 1-shallow center and there are at least two points among $a_1,\dots,a_g$ being sampled to $S'$, and $\tau = 0$ otherwise.
Observe that $\sigma(c) \geq \tau$.
Thus, we have
\begin{equation*}
\mathbf{E}[\sigma(c)] \geq \mathbf{E}[\tau] = \left( 1- \frac{1}{d} \right)^\theta \left( \frac{g}{d} - \frac{g}{d} \left( 1- \frac{1}{d} \right)^{g-1} \right) \geq \frac{g}{4d^2},
\end{equation*}
since $\theta<d$ and $g \geq 2$.
It follows that
\begin{equation*}
\frac{1}{4d^2} \sum_{c \in H_d} \mathit{deg}(c) \leq \sum_{c \in H_d} \mathbf{E}[\sigma(c)] \leq \mathbf{E}[\varphi] \leq \frac{2n}{d},
\end{equation*}
where $H_d$ is the set of the $d$-shallow centers of $S$.
As a result, the degree-sum of the $d$-shallow centers of $S$ is at most $8dn$, completing the proof.
\subsection{Proof of Lemma~\ref{okshallow}}
Suppose $\pi_{a_1},\dots,\pi_{a_n} \in [\varepsilon,1]$ for a constant $\varepsilon>0$.
Let $c$ be a critical center of $S$ with $\mathit{dep}_S(c) = d$.
Without loss of generality, we assume \\
$\bullet$ $\mathit{dist}(a_1,\hat{c}) \leq \mathit{dist}(a_2,\hat{c}) \leq \dots \leq \mathit{dist}(a_d,\hat{c}) < \mathit{diam}(c)$, \\
$\bullet$ $a_{d+1}$ is involved by $c$ and in the $k$-LNN of $\hat{c}$. \\
We claim that $d=O(k)$.
The claim is trivial when $d \leq k$, thus assume $d>k$.
Since $a_{d+1}$ is in the $k$-LNN of $\hat{c}$, there must exist $i \leq k$ such that $\mathit{NNP}_{\hat{c}}(a_i) < \mathit{NNP}_{\hat{c}}(a_{d+1})$.
It then follows that
\begin{equation*}
(1-\varepsilon)^{d-i+1} \geq \prod_{j=i}^{d} (1-\pi_{a_j}) \geq \pi_{a_{d+1}} \prod_{j=i}^{d} (1-\pi_{a_j}) > \pi_{a_i} \geq \varepsilon.
\end{equation*}
As a result, $d < \log_{1-\varepsilon} \varepsilon + i-1 \leq \log_{1-\varepsilon} \varepsilon + k = O(k)$.
\subsection{Proof of Lemma~\ref{condforcrit}}
Without loss of generality, assume $a_1,\dots,a_k$ are $k$ points closer to $\hat{c}$ than the $c$-involved points and $\pi_{a_1},\dots,\pi_{a_k}$ are greater than $\mu_0$.
Let $x \in S$ be any point involved by $c$.
Since $\pi_x$ is drawn from $f$, we must have $\pi_x \leq \mu$ by definition.
We now show that $x$ is not in the $k$-LNN of $\hat{c}$.
We have the inequality
\begin{equation*}
\frac{\mathit{NNP}_{\hat{c}}(x)}{\mathit{NNP}_{\hat{c}}(a_i)} \leq \frac{\pi_x (1-\pi_{a_i})}{\pi_{a_i}} < \frac{\mu (1-\mu_0)}{\mu_0} = 1,
\end{equation*}
for $i \in \{1,\dots,k\}$.
It follows that there are at least $k$ points in $S$ which have greater probabilities of being the nearest-neighbor of $\hat{c}$ than $x$.
Thus, $x$ is not in the $k$-LNN of $\hat{c}$.
Since $x$ is arbitrarily chosen, we know that $c$ is not critical, which completes the proof.

\section{A remark about the spatial assumption} \label{remark}
In many geometric problems, it is usually reasonable to make some general position assumptions about the data points for convenience of proof and exposition.
The reason is that one may handle the degenerate cases by applying a small perturbation to the data points.
Many geometric properties of the dataset are insensitive to such a small perturbation.
For instance, the Voronoi diagrams in Euclidean spaces, or more generally, in uniformly convex normed spaces, are known to be stable under a small perturbation of the sites \cite{reem2011geometric}.
However, for tree-space LVD, this is not the case.
A small perturbation of the stochastic points in $\mathcal{T}$ may significantly influence the tree-space LVD.
A very simple example is presented in the following figure.
As we see, if we slightly perturb $a_3$, even the structure of the 1-LVD changes significantly.
Therefore, when studying LVD and LNN search in tree spaces, it is not natural to make spatial assumptions about the given stochastic points as well as their midpoints.
\begin{center}
    \includegraphics[height=4.5cm]{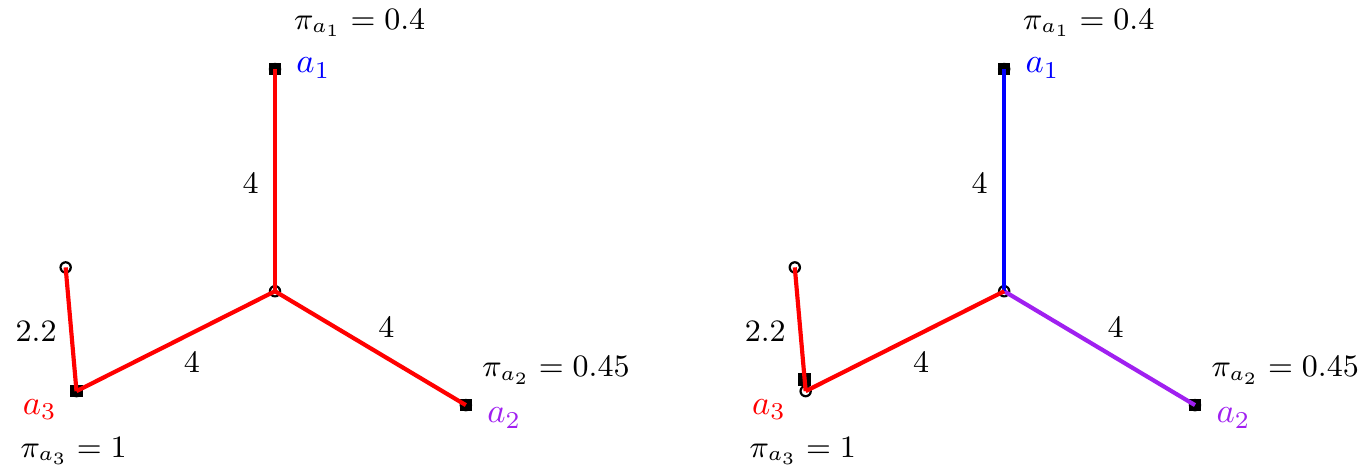}
\end{center}

\section{Details for constructing LVD data structure} \label{detconstr}
\subsection{Computing and sorting the centers} \label{compmid}
First of all, we apply Theorem~\ref{preproc} to obtain a new tree-space
    $\mathcal{T}'$ represented by an $O(n)$-vertex tree $T'$ such that
    $S \subset \mathcal{T}' \subseteq \mathcal{T}$ and each stochastic point in
    $S$ is located at a vertex of $T'$.
This step takes $O(t + n \log n)$ time.
Note that all the centers of $S$ must be in $\mathcal{T}'$, so we can first
    work on $\mathcal{T}'$ and then map the computed centers back to
    $\mathcal{T}$.
Before computing the centers, we do some preprocessing on the tree $T'$.
For all pairs $(e,v)$ where $e$ is an edge and $v$ is a vertex of $T'$,
    we figure out the side of $e$ that $v$ locates on.
This can be easily done in $O(n^2)$ time with a careful implementation.
Furthermore, for each vertex $v$ of $T'$, we create a sorted list $B_v$
    which contains all points in $S$ sorted according to their distances to $v$.
This step can also be done in $O(n^2)$ time as follows.
Observe that, if $v$ and $v'$ are adjacent vertices connected by an edge $e$,
    we can modify the sorted list $B_v$ to obtain the list $B_{v'}$.
Specifically, we separate $B_v$ into two sorted sublists each of which contains
    the stochastic points on one side of $e$.
Then $B_{v'}$ can be computed by merging these two sorted sublists in
    $O(n)$ time.
Based on this observation, we can first straightforwardly create the sorted
    list for one vertex of $T'$ in $O(n \log n)$ time, and keep modifying it to
    obtain the lists for other vertices, which takes $O(n^2)$ time in total.
After the preprocessing, we are ready to compute the centers of $S$.
The centers lying at any vertex $v$ of $T'$ can be directly found from the
    sorted list $B_v$.
To compute the centers lying in the interior of an edge $e = (v,v')$,
    we utilize the sorted list $B_v$ (or $B_{v'}$).
Again, we separate $B_v$ into two sorted sublists (say $B_v'$ and $B_v''$)
    each of which contains the stochastic points on one side of $e$.
We notice that a center in the interior of $e$ involves a set $A'$ of
    stochastic points located at the vertices in $B_v'$ and a set $A''$ of
    stochastic points located at the vertices in $B_v''$.
The points in $A'$ must have the same distance to $v$ (say $d'$), so are the
    points in $A''$ (say $d''$).
Furthermore, we must have $0<d''-d'<w$, where $w$ is the weight (length) of $e$.
With these observations, one can easily apply the standard sliding window
    technique to compute the centers in the interior of $e$ in $O(\alpha+n)$ time where $\alpha$ is the number of the centers computed.
Thus, the computation for all edges takes $O(n^2)$ time.
After the centers are computed, we sort the centers in the interior of each
    edge $e$ in the order they appear on $e$.
This part takes $O(n^2 \log n)$ time in worst case.
The final step is to map the centers back to the original tree
    space $\mathcal{T}$.
If $\mathcal{T}'$ is constructed by applying the method in
    Appendix~\ref{preproc}, then it is easy to keep a ``relation''
    between $T'$ and $T$ during the construction.
For example, for each edge $e$ of $T'$, we can record the edges of $T$
    intersecting with $e$ in the order the intersections appear on $e$.
With this information, as long as the centers in the interior of each edge of
    $T'$ is sorted, the entire mapping process can be done in $O(t+n^2)$ time.
At the end, after we map the centers to $\mathcal{T}$, we need to do another
    sort for the centers in the interior of each edge of $T$.
The overall time cost for computing and sorting the centers is
    $O(t+n^2 \log n)$.

\subsection{Constructing the LVD in the walk} \label{walk}
During the walk, the nearest-neighbor probabilities of $a_1,\dots,a_n$ change
    only when we arrive at (resp., leave from) a center $c$ from (resp., to)
    one of its branches.
At this time, we need to update the nearest-neighbor probabilities, maintain
    the binary search tree, and (possibly) retrieve the $k$-LNN from the binary
    search tree.
Let $m_c$ be the number of the stochastic points involved by $c$.
Note that only these $m_c$ stochastic points may change their nearest-neighbor
    probabilities (this may be not true if there are other centers which have
    the same location as $c$, but the changes of the nearest-neighbor
    probabilities of the points involved by other centers can be charged to
    those centers instead of $c$).
The update of the nearest-neighbor probabilities can be easily done in $O(m_c)$
    time, if we store (before the walk) for each branch of a center $c$ the
    product of the non-existence probabilities of the $c$-involved points in
    this branch.
The maintenance of the binary search tree is achieved by $O(m_c)$ deletion and
    insertion operations, and thus takes $O(m_c \log n)$ time.
Finally, the time for retrieving the $k$-LNN from the binary search tree is
    $O(\log n+k)$.
Therefore, at every time we arrive at (resp., leave from) $c$ from (resp., to)
    one of its branches in the walk, we spend $O(m_c \log n+k)$ time.
During the walk, we arrive at (resp., leave from) $c$ from (resp., to) its
    branches $O(\mathit{deg}(c))$ times in total.
It follows that the time cost charged to $c$ is
    $O(\mathit{deg}(c)\cdot m_c \log n + \mathit{deg}(c) \cdot k)$.
Since we have $\sum_c \mathit{deg}(c)\cdot m_c = O(n^2)$, the overall time cost
    for the walk is $O(t + n^2 \log n + n^2 k)$.
(There are also some low-level details for implementing the walk, e.g., how to
    know whether we are arriving at a center from one of its branches, etc.
Such issues can be easily handled with enough preprocessing work before the
    walk.)

\end{document}